\documentclass[twocolumn]{aastex631}

\usepackage{amssymb,amsmath}
\usepackage{graphicx}
\usepackage{natbib}

\usepackage{multirow}

\shorttitle{}
\shortauthors{Jeffrey, Krucker et al.}

\begin{document}

\title{A Modelling Investigation for Solar Flare X-ray Stereoscopy with Solar Orbiter/STIX and Earth Orbiting Missions}

\author[0000-0001-6583-1989]{Natasha L. S. Jeffrey}
\affiliation{Department of Mathematics, Physics \& Electrical Engineering, Northumbria University,
Newcastle upon Tyne, UK, NE1 8ST}

\author{S\"{a}m Krucker}
\affiliation{Institute for Data Science, University of Applied Sciences and Arts Northwestern Switzerland (FHNW), Bahnhofstrasse 6, 5210 Windisch, Switzerland}
\affiliation{Space Sciences Laboratory, University of California, 7 Gauss Way, 94720 Berkeley, USA}

\author[0000-0002-6060-8048]{Morgan Stores}
\affiliation{Department of Mathematics, Physics \& Electrical Engineering, Northumbria University,
Newcastle upon Tyne, UK, NE1 8ST}

\author[0000-0002-8078-0902]{Eduard P. Kontar}
\affiliation{School of Physics \& Astronomy, University of Glasgow, Glasgow, G12 8QQ, UK}

\author{Pascal Saint-Hilaire}
\affiliation{Space Sciences Laboratory, University of California, 7 Gauss Way, 94720 Berkeley, USA}

\author{Andrea F. Battaglia}
\affiliation{Institute for Data Science, University of Applied Sciences and Arts Northwestern Switzerland (FHNW), Bahnhofstrasse 6, 5210 Windisch, Switzerland}
\affiliation{ETH Z\"{u}rich, R\"{a}mistrasse 101, 8092 Z\"{u}rich Switzerland}

\author{Laura Hayes}
\affiliation{European Space Agency (ESA), European Space Research and Technology Centre (ESTEC), Keplerlaan 1, NL-2201 AZ Noordwijk, the Netherlands}

\author{Hannah Collier}
\affiliation{Institute for Data Science, University of Applied Sciences and Arts Northwestern Switzerland (FHNW), Bahnhofstrasse 6, 5210 Windisch, Switzerland}
\affiliation{ETH Z\"{u}rich, R\"{a}mistrasse 101, 8092 Z\"{u}rich Switzerland}

\author{Astrid Veronig}
\affiliation{University of Graz, Institute of Physics \& Kanzelh\"{o}he Observatory for Solar and Environmental Research, Kanzelh\"{o}he 19, 9521 Treffen, Austria}

\author{Yang Su}
\affiliation{Purple Mountain Observatory, Chinese Academy of Sciences, Nanjing, China}

\author{Srikar Paavan Tadepalli}
\affiliation{ISRO UR Rao Satellite Center, Vimanapura, Bangalore, Karnataka - 560017, India}

\author{Fanxiaoyu Xia}
\affiliation{Purple Mountain Observatory, Chinese Academy of Sciences, Nanjing, China}

\begin{abstract}
The Spectrometer/Telescope for Imaging X-rays (STIX) on board Solar Orbiter (SolO) provides a unique opportunity to systematically perform stereoscopic X-ray observations of solar flares with current and upcoming X-ray missions at Earth. These observations will produce the first reliable measurements of hard X-ray (HXR) directivity in decades, providing a new diagnostic of the flare-accelerated electron angular distribution and helping to constrain the processes that accelerate electrons in flares. However, such observations must be compared to modelling, taking into account electron and X-ray transport effects and realistic plasma conditions, all of which can change the properties of the measured HXR directivity. Here, we show how HXR directivity, defined as the ratio of X-ray spectra at different spacecraft viewing angles, varies with different electron and flare properties (e.g., electron angular distribution, highest energy electrons, and magnetic configuration), and how modelling can be used to extract these typically unknown properties from the data. Lastly, we present a preliminary HXR directivity analysis of two flares, observed by the Fermi Gamma-ray Burst Monitor (GBM) and SolO/STIX, demonstrating the feasibility and challenges associated with such observations, and how HXR directivity can be extracted by comparison with the modelling presented here. 
\end{abstract}

\section{Introduction}
Solar flares are a product of magnetic reconnection \citep[e.g.,][]{parker1957sweet,sweet195814,priest2002magnetic,2013NatPh...9..489S} in the Sun's atmosphere and the release and conversion of magnetic energy into other energies including accelerating non-thermal charged particles \citep[e.g.,][]{2008LRSP....5....1B}. Observations deduce that flares are exceptionally efficient particle accelerators with 10-50\% of the released magnetic energy going into particle acceleration \citep[e.g., ][]{emslie2012global,warmuth2016constraints,2017ApJ...836...17A}. However, the exact mechanisms and locations of energy release and/or acceleration are not well-constrained. Competing theories of acceleration include magnetic energy dissipation by plasma waves and turbulence \citep{larosa1993mechanism,1996ApJ...461..445M,2016ApJ...827L...3V,petrosian2012stochastic,kontar2017turbulent} and/or plasma instabilities and the formation of magnetic islands \citep{2006Natur.443..553D,2011NatPh...7..539D,2014PhPl...21i2304D} and/or shock acceleration \citep[e.g.,][]{2015Sci...350.1238C}. 

\begin{figure*}[hbpt!]
\centering
\includegraphics[width=0.99\textwidth]{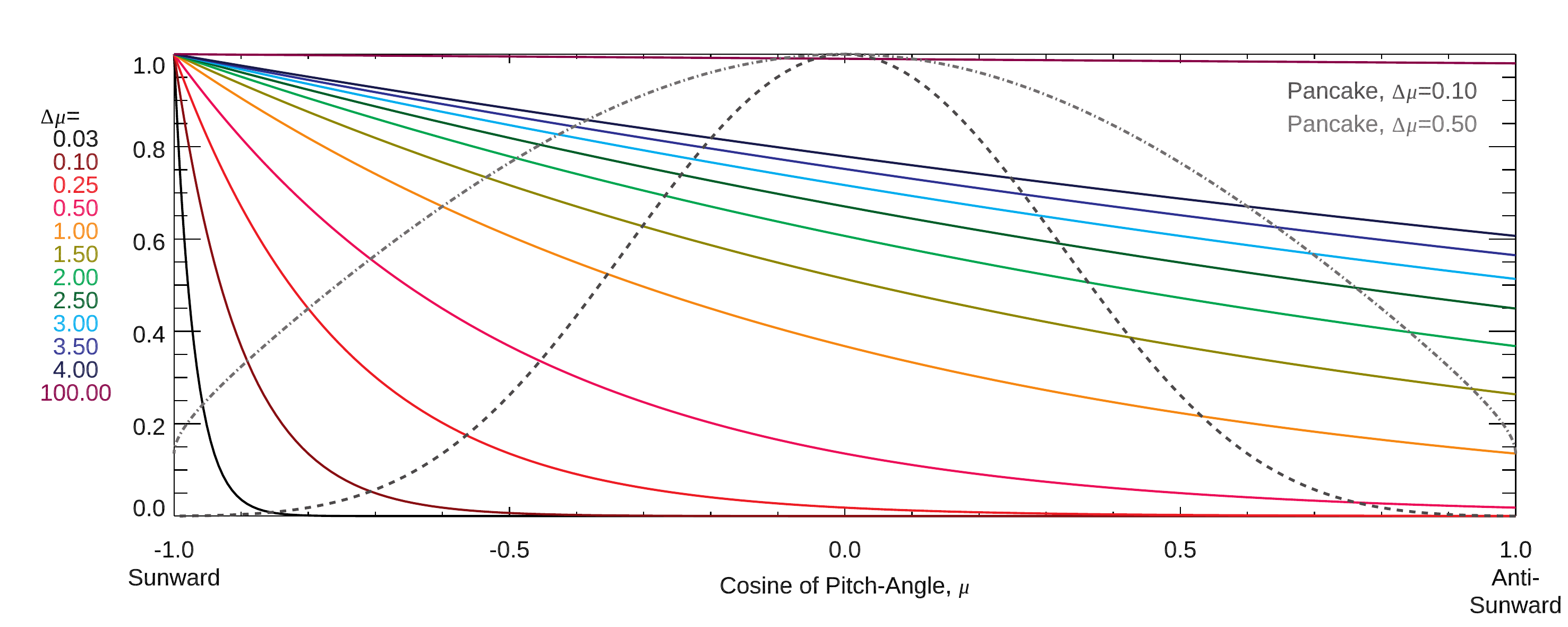}
\caption{Examples of beamed to isotropic electron directivity (pitch-angle distribution) controlled by the parameter $\Delta\mu$ in Equation \ref{td_mu}. Here, small values of $\Delta\mu$ produce highly beamed distributions with the bulk of the electron populations moving sunward while large values of $\Delta\mu$ produce closer to isotropic distributions. The grey dashed and dashed-dot lines also show two ``pancake'' distribution examples (Equation \ref{td_mu_2} with $\Delta\mu=0.1$ and $\Delta\mu=0.5$) where a high fraction of the electrons are directed at an angle of $90^{\circ}$ to the guiding magnetic field, resulting in a pitch-angle distribution that may be more suitable for certain mechanisms.}
\label{Fig0}
\end{figure*}

In a standard flare model, electrons are accelerated along newly formed, closed magnetic field lines and precipitate into the dense layers of the lower atmosphere where they lose energy producing hard X-rays (HXRs) \citep[\textbf{cf.}][]{holman2011implications}, while other electrons may escape into the heliosphere on open field lines, as solar energetic electrons (SEE) \citep[e.g., ][]{2007ApJ...663L.109K,klein2017acceleration}. X-rays are a direct link to flare-accelerated electrons at the Sun and a vital probe of the physical processes occurring in flares \citep{2003ApJ...595L.115B,kontar2011deducing}. Over the last twenty years, the flare X-ray energy spectrum has been well observed by instruments such as the Ramaty High Energy Solar Stereoscopic Imager (RHESSI; \citealt{2002SoPh..210....3L}). However, many of the important properties required to constrain the acceleration process(es) still remain elusive, since they are difficult to determine from an X-ray spectrum viewed by a single spacecraft alone. One important property, the HXR directivity, is a vital diagnostic of the emitting, and the accelerated, electron pitch-angle distribution (often used interchangeably with electron anisotropy or directivity). The electron pitch-angle is defined as the direction of electron velocity with respect to the guiding magnetic field and thus, is a key diagnostic of the dominant acceleration mechanism. As one example, accelerated electrons produced by a stochastic acceleration mechanism with efficient scattering (short timescales) are expected to produce isotropic pitch-angle distributions (although, isotropy is usually prescribed in such models to increase acceleration efficiency) \citep[e.g.,][]{1994ApJS...90..623M,1996ApJ...461..445M,petrosian2012stochastic}. Moreover, processes such as resonant scattering due to waves propagating parallel to the magnetic field may preferentially scatter at $90^{\circ}$ (e.g., \citealt{1999ApJ...527..945P}).  In order to understand HXR directivity, we must also take into account electron transport effects in the solar plasma that broaden the electron distribution, increasing the isotropy by e.g., collisional \citep[][]{jeffrey2014spatial,2015ApJ...809...35K,2019ApJ...871..225K} and/or non-collisional pitch-angle scattering \citep{2014ApJ...780..176K,2018A&A...610A...6M}. Thus, even if the accelerated distribution is strongly beamed, the angular distribution of radiating electrons is expected to isotropise as they are transported from the corona to the chromosphere, where the bulk of HXRs are produced and observed due to the bremsstrahlung dependency on density.

\begin{figure*}[hbpt!]
\centering
\includegraphics[width=0.69\textwidth,angle=270]{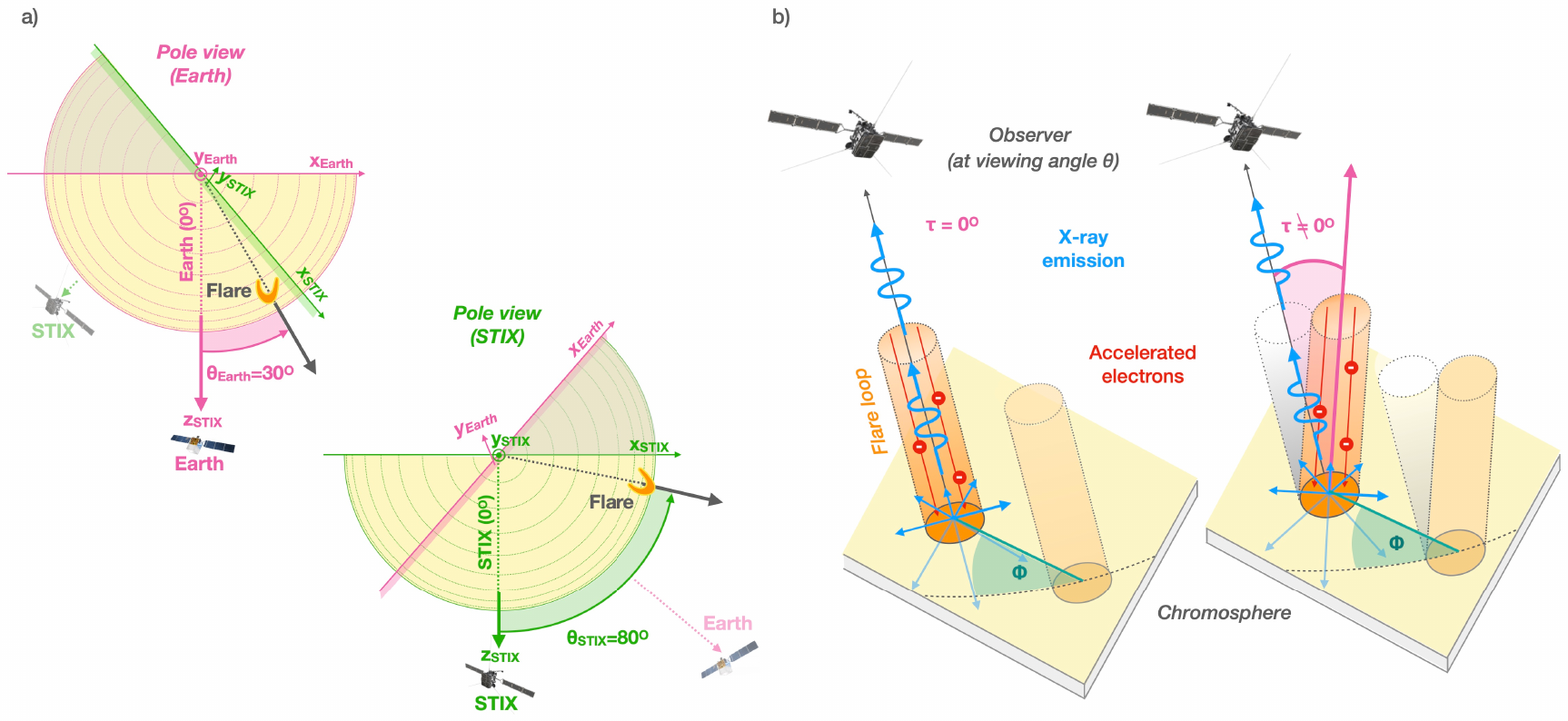}
\caption{a) Example of a flare, observed by two different spacecraft, at two different heliocentric (viewing) angles, $\theta_{\rm Earth}$ viewed by spacecraft at Earth and $\theta_{\rm STIX}$ viewed by STIX. Locally, the heliocentric angle is measured in the plane of the spacecraft, flare and the solar centre (which can differ for each spacecraft), given by $\theta=\arcsin(\sqrt{x^2 + y^2}/R_{\odot})$ i.e., the $x$ and $y$ axes are defined relative to the observer position, as shown. b) Locally, the flare can also be tilted with respect to the local solar radial direction \citep{2008ApJ...674..570E}. This tilt is represented by $\tau$ where $\tau=0^{\circ}$ represents a scenario where the flaring loop apex is aligned with local solar radial direction. The effect of $\tau$ on HXR directivity is also dependent on $\Phi$, the azimuthal direction of the axis around which the loop is tilted \citep{2008ApJ...674..570E}.}
\label{fig_1}
\end{figure*}
Due to different issues, HXR directivity has been difficult to routinely measure to date. We have sought to measure the HXR and electron directivity in several different ways: (1) HXR albedo mirror analysis of strong solar flares \citep[][]{1978ApJ...219..705B,2006ApJ...653L.149K}. Novel, but limited albedo mirror (X-ray Compton backscattering in the photosphere) analyses suggests that the HXR emitting electron distribution is close to isotropic especially below 150 keV, at least for the few events published \citep[][]{2013SoPh..284..405D}. 
(2) Using statistical flare studies of centre-to-limb variations in e.g., flux or spectral index. Statistical studies such as \citet{2007A&A...466..705K} studied 398 flares but gave no clear conclusion regarding average flare directivity, particularly because the study was only able to investigate in the (mainly thermal) $15-20$~keV range.
(3) Using linear X-ray polarization measurements from a single flare with one satellite \citep[e.g.,][]{1970SoPh...14..204T,2004AdSpR..34..462M}. There is a direct link between X-ray linear polarization and electron anisotropy \citep[][]{1983ApJ...269..715L,1978ApJ...219..705B,2011A&A...536A..93J,2020A&A...642A..79J}. Nevertheless, observations with past instruments and non-dedicated polarimeters (e.g., RHESSI), have proved problematic, owing to instrumental issues, although rare RHESSI observations suggested some level of directivity e.g., \citet{2006SoPh..239..149S}. Currently, there is no solar-dedicated X-ray polarimeter to measure directivity, but the PolArization and Directivity X-Ray Experiment (PADRE; a CubeSat planned to be launched in 2025) will be capable of X-ray spectro-polarimetry up to $\approx 100$~keV.
(4) By simultaneously observing a single flare with two satellites at different viewing angles (i.e., measuring HXR directivity). Previous stereoscopic studies \citep[e.g., ][]{1992AAS...180.2307K,1998ApJ...500.1003K} found no clear evidence for directivity at X-ray energies between $25-125$~keV. However, such observations can suffer from calibration issues, making the results unreliable. Thus, it is fundamental that the two instruments have a well-known energy cross-calibration before any data is interpreted. Further, previous stereoscopic studies did not take effects such as X-ray albedo into account.
Now with the successful deployment of Solar Orbiter (SolO) \citep{2020A&A...642A...1M} and its Spectrometer/Telescope for Imaging X-rays (STIX) \citep{2020A&A...642A..15K} we are now able to detect flares with different viewing angles from the Earth-Sun line as close as 0.28 AU (at perihelion) and up to inclinations of $\approx25^{\circ}$. Here, we perform modelling that can be compared directly with stereoscopic observations from SolO/STIX and current or near-future Earth orbiting missions such as the Gamma-ray Burst Monitor (GBM) onboard Fermi \citep{2009ApJ...702..791M}, Advanced Space-based Solar Observatory/Hard X-ray Imager (ASO-S/HXI) \citep{2019RAA....19..160Z,2019RAA....19..167K,2019RAA....19..163S}, Aditya-HEL1OS, PADRE, and beyond, acting as a foundation for extracting the electron directivity and other electron and flare properties from stereoscopic observations. 
Section \ref{model} provides an overview of the preliminary modelling, Section \ref{results} displays and discusses a selection of modelling results showing how certain electron and flare parameters can be extracted from HXR stereoscopy, while a preliminary analysis of HXR directivity from two flares observed by SolO/STIX and Fermi/GBM is shown in Section \ref{prem}. Section \ref{summ} summarizes the main results of this study.

\begin{figure*}[hbtp!]
\centering
\includegraphics[width=0.49\textwidth]{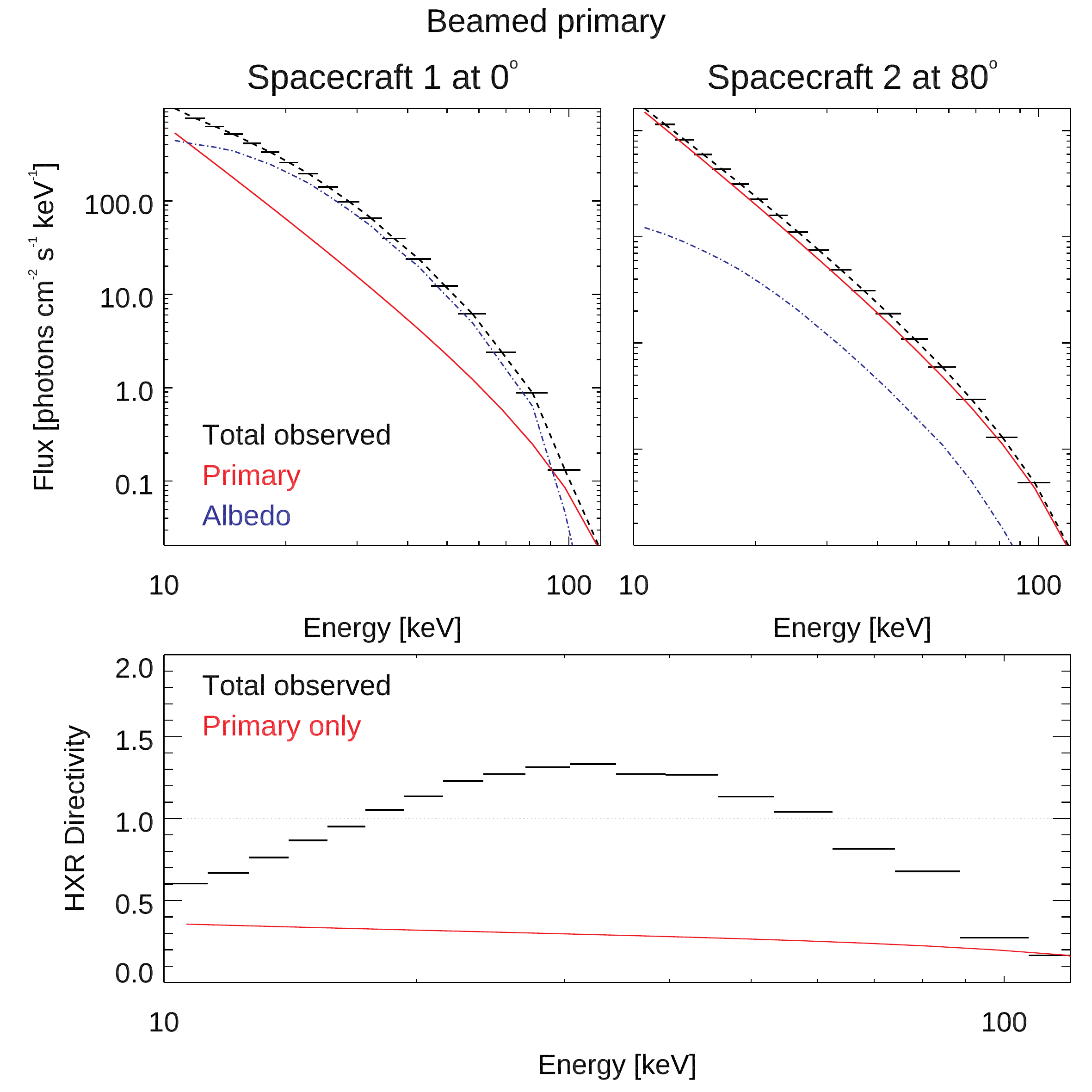}
\includegraphics[width=0.49\textwidth]{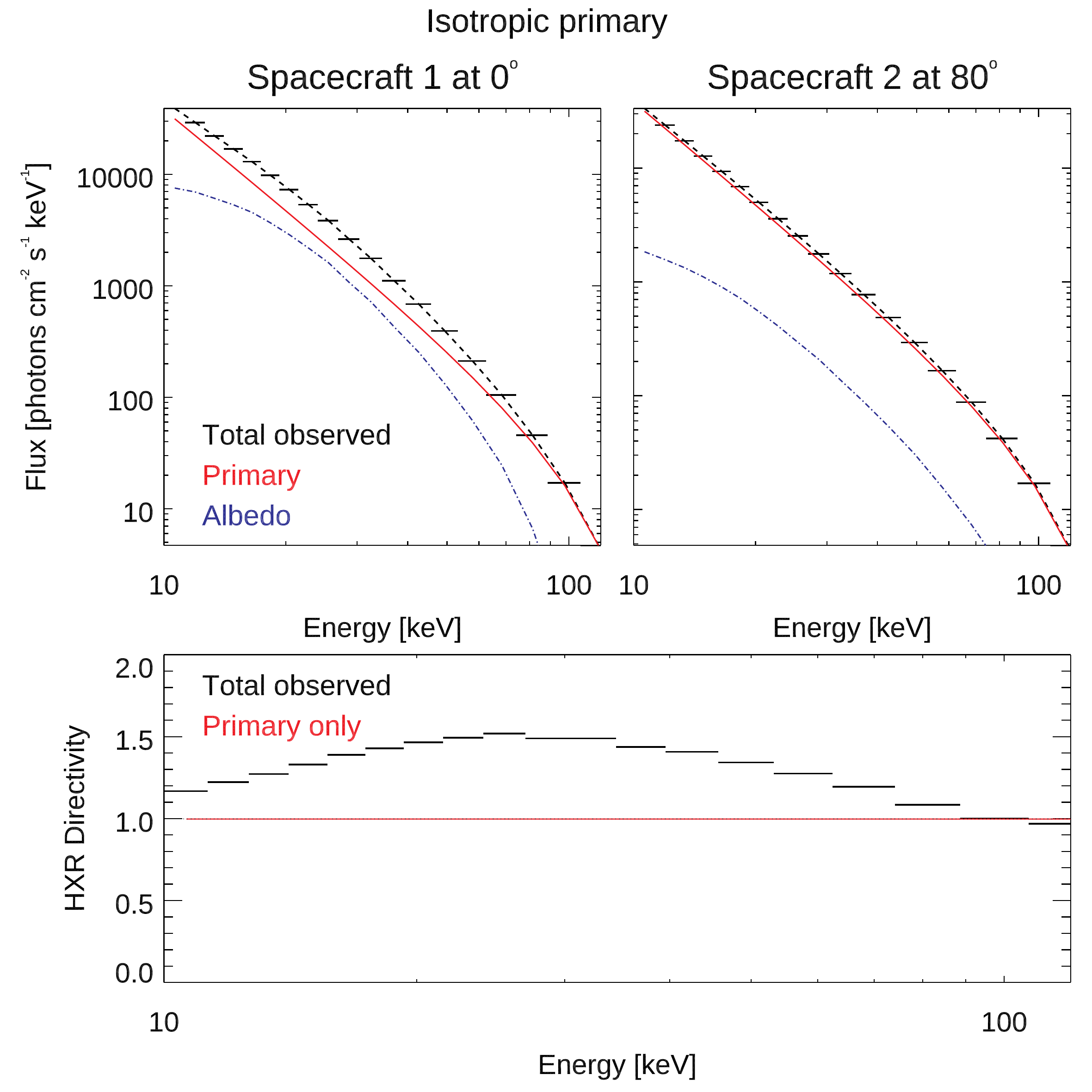}
\caption{Top row: examples of model spectra (black: total observed X-ray emission) and individual components (red: primary X-ray emission and blue: albedo X-ray emission) for a flare with a beamed (towards the solar surface) electron directivity (left) and an isotropic electron directivity (right). In each case, the spectra are observed at two different spacecraft viewing angles of $0^{\circ}$ (disk centre) and $80^{\circ}$ (close to the limb). Bottom row: resulting HXR directivity (black: total observed and red: primary emission only) found from the ratio of spectra for the beamed (left) and isotropic (right) cases. These examples show the importance of accounting for the X-ray albedo component, which is greatest at the disk centre and smallest at the limb, while increasing for greater downward directivity.}
\label{fig_00}
\end{figure*}

\section{Models of HXR directivity}\label{model}

\subsection{Coronal transport-dependent modelling}
Following previous works \citep{2014ApJ...787...86J,2015ApJ...809...35K,2019ApJ...880..136J,2023ApJ...946...53S}, the evolution of an electron flux $F$ in energy $E$ [erg], cosine of the pitch-angle ($\beta$), $\mu=\cos\beta$, and distance along the guiding magnetic field $z$ [cm], from a coronal loop apex to the chromosphere can be modelled using a time-independent Fokker-Planck equation (see Equation \ref{time_independent_Fokker_planck}). This takes into account the processes that alter electron properties including their directivity, e.g., Coulomb collisions (represented by the last two terms on the right-hand side (RHS) of Equation \ref{time_independent_Fokker_planck}) and turbulent scattering (represented by the first term on the RHS of Equation \ref{time_independent_Fokker_planck}) from magnetic fluctuations using a diffusion coefficient $D_{\mu\mu}$, often chosen to be isotropic for simplicity, i.e., \citet{2020A&A...642A..79J}.

\begin{equation}
\begin{split}
        \mu \frac{\partial F}{\partial z} & =       \underbrace{\sqrt{\frac{m_{e}}{2E}}\bigg \{\frac{\partial}{\partial \mu}\left[D_{\mu\mu}(\mu,z)\frac{\partial F}{\partial \mu}\right]\bigg \}}_\text{turbulent scattering} \\
        &+ \underbrace{\Gamma m_e^2 \bigg \{ \frac{\partial}{\partial E} \bigg [ G(u[E]) \frac{\partial F}{\partial E} + \frac{G(u[E])}{E}\bigg ( \frac{E}{k_B T} - 1\bigg ) F \bigg ]}_\text{collisional energy losses} \bigg \} \\
        & + \underbrace{\frac{\Gamma m_e^2}{8E^2} \bigg \{\frac{\partial}{\partial \mu} \bigg [ (1 - \mu^2) \left [ \text{erf}(u[E]) - G(u[E])\right ]\frac{\partial F }{\partial \mu} \bigg ] \bigg \}}_\text{collisional pitch-angle scattering}
        \end{split}
        \label{time_independent_Fokker_planck}
\end{equation}
 where $\Gamma = 4 \pi e^4$ln$\Lambda n / m_e^2$, for electron charge $e$ [statC], Coulomb logarithm ln$\Lambda$, electron mass $m_e$ [g] and coronal number density $n$ [cm$^{-3}$]. The error function is given by erf$(u)$ and $G(u)=(\text{erf}(u) - u \, \text{erf}\,'(u))/2u^2$
where $u$ is the dimensionless velocity $u = v/(\sqrt{2} v_{th})$, $v$ is the velocity [cm s$^{-1}$], $v_{th} = \sqrt{k_BT/m_e}$ [cm s$^{-1}$] for Boltzmann constant $k_{B}$ and coronal temperature $T$, and $\text{erf}\,'(u)=\frac{d\text{erf}}{du}$. Such functions control the lower-energy ($E\approx k_{B}T$) electron interactions ensuring that they become indistinguishable from the background thermal plasma. 

To allow the evolution of an electron distribution to be modelled in space, energy, and pitch-angle to the guiding magnetic field, Equation \ref{time_independent_Fokker_planck} can be solved numerically by its conversion into a set of time-independent stochastic differential equations (SDEs) \citep[e.g.,][]{1986ApOpt..25.3145G,2017SSRv..212..151S} for $E$, $\mu$ and $z$. 

\begin{figure*}[hbpt!]
    \centering
    \small{{\sffamily \textcolor{darkgray}{Dependence on Spectral Index}}}\\
        \includegraphics[width=0.48\textwidth]{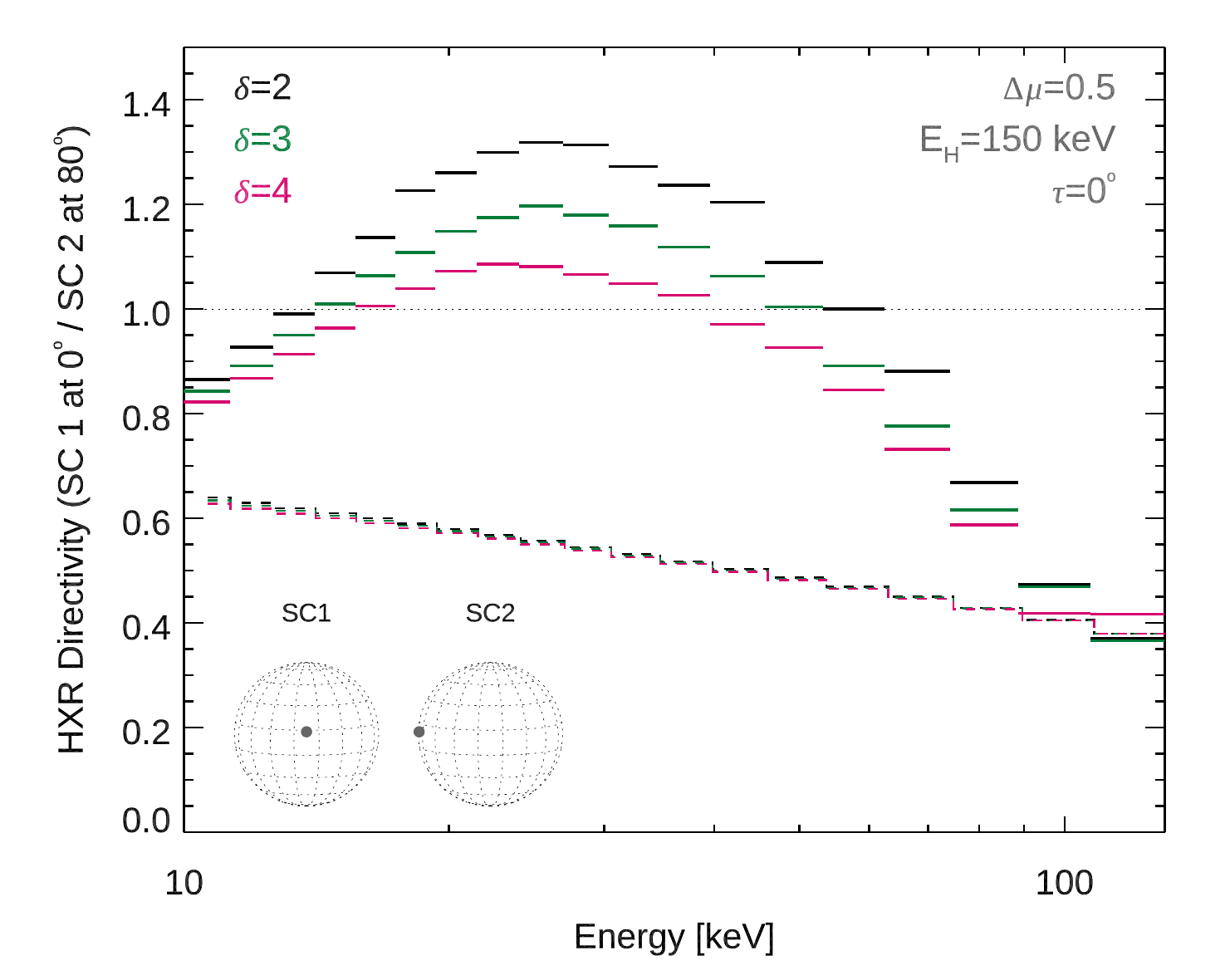}
    \includegraphics[width=0.48\textwidth]{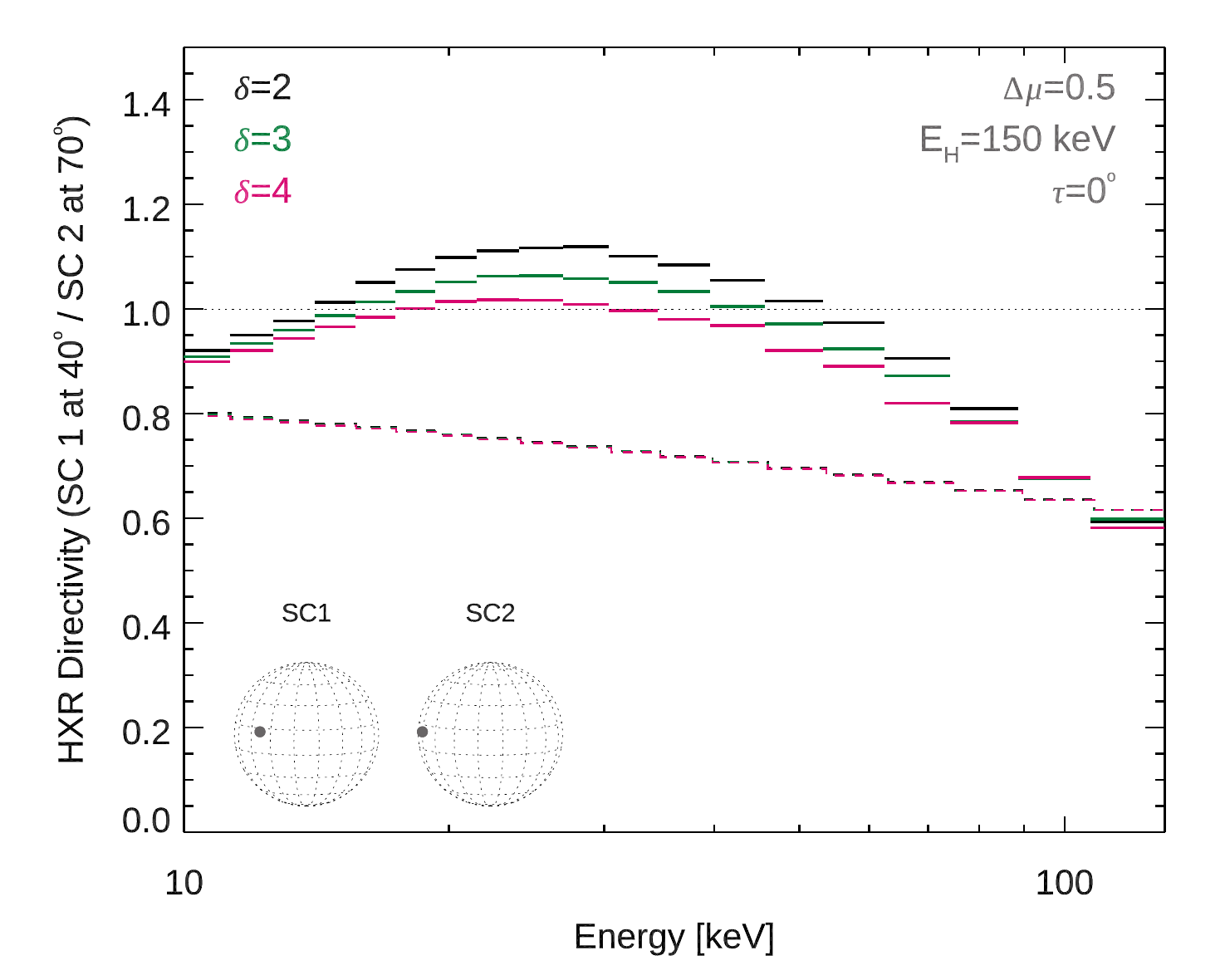}
    \caption{HXR directivity (ratio of HXR spectra) for spacecraft viewing angles of $0^{\circ}$ and $80^{\circ}$ (left) and $40^{\circ}$ and $70^{\circ}$ (right), showing how the directivity can change with the power law spectral index of the electron spectrum (using $\delta\equiv\delta_{FP}=2,3,4$ (footpoint)). In these examples, we use an electron anisotropy of $\Delta\mu=0.5$, an electron high energy cutoff of $E_{H}=150$~keV, and a loop tilt of $\tau=0^{\circ}$. Solid bars: total observed flux ratio, dashed lines: primary component only, without albedo.}
    \label{fig2}
\end{figure*}

\subsection{Transport-independent modelling}
For a simple comparison and for a basic understanding of the expected results, we also perform transport-dependent simulations (similar to \citet{2011A&A...536A..93J}), where a chosen electron flux $F(E,\mu)$ is deposited into the chromosphere, and modelled using,

\begin{equation}\label{td_mu}
F(E,\mu)\propto E^{-\delta_{\rm FP}}  \exp\left(-\frac{(1+\mu)}{\Delta \mu}\right).
\end{equation}

This function uses a single parameter $\Delta\mu$ to alter the electron directivity from beamed to isotropic, where large values of $\Delta\mu$ ($>>1$) produce isotropic distributions and small values of $\Delta\mu$ ($<<1)$ produce field-aligned `beamed' distributions (Figure \ref{Fig0}). Here, the electron spectral index $\delta_{FP}$ is the spectral index related to the footpoint emission assuming a thick-target chromosphere (not the injected or accelerated power law index). Again, for this parameter study, we assume the electron properties remain identical at all emitted pitch-angles $\mu = \cos \beta$.

It is also interesting to consider a `pancake' directivity (where the bulk of the emitting electrons have their velocity directed at $90^{\circ}$ to the guiding magnetic field, see Figure \ref{Fig0}) using
\begin{equation}\label{td_mu_2}
F(E,\mu)\propto E^{-\delta_{\rm FP}} \exp\left(-\frac{(1-\sqrt{1-\mu^{2}})}{\Delta \mu}\right).
\end{equation}
While solar models usually invoke isotropic turbulence as accelerator and/or scatterer (since the form of the turbulence is unconstrained), solar wind studies, guided by in-situ measurement, show anisotropy with field fluctuations greater in the direction perpendicular to the field (i.e., $\delta B_{\perp}>\delta B_{||}$) e.g., \citet{2013SSRv..178..101A}.

In all modelling scenarios, the resulting X-ray bremsstrahlung distribution is calculated using the full polarization angle-dependent bremsstrahlung cross section \citep{1953PhRv...90.1030G,1972SoPh...25..425H}, using the forms of  \citet{2008ApJ...674..570E} and \citet{2011A&A...536A..93J}.

Finally, the photospheric `backscattered' X-ray albedo component, which is critical for the correct determination of the HXR directivity (Figure \ref{fig_00}), is modelled using the Monte Carlo albedo code of \citet{2011A&A...536A..93J}. The albedo X-rays are viewed alongside those X-rays directly emitted from the HXR source, which are often called the primary X-rays. The albedo component consists of energy and pitch-angle altered photons and creates a `bump’ in the X-ray spectrum over the energies of 10-100 keV with a peak around the 20-50 keV range (see Figure \ref{fig_00}) An isotropic HXR source produces the minimum albedo and even its flux can account for up to 40\% of the detected flux in the peak albedo energy range between 20 and 50 keV \citep[e.g., ][]{1978ApJ...219..705B,2006ApJ...653L.149K,2007A&A...466..705K}.

The full X-ray distribution (direct primary emission from the chromosphere plus the photospheric albedo component) is examined for various spacecraft viewing angles. We do not perform an exhaustive parameter study here and set the height of all chromospheric emission to $h=1$~Mm when calculating the X-ray albedo component. 

\subsection{Measuring the HXR directivity}

Any flare location on the solar disk (or the flare viewing angle) can be defined by its radial position given by $\sqrt{x^{2}+y^{2}}$ using the local flare ($x$,$y$) coordinates (with the corresponding solar centre viewed by each spacecraft defined as [$x=0$, $y=0$].) The heliocentric angle is measured in the plane of the spacecraft, flare and the solar centre (this plane can differ for each spacecraft), given by $\theta=\arcsin(\sqrt{x^2 + y^2}/R_{\odot})$ where $R_{\odot}$ is the solar radius for each spacecraft observation. An example of a flare observed at two heliocentric angles by two different spacecraft is shown in Figure \ref{fig_1}, panel a).

In all simulations (similar to X-ray observations), a heliocentric angle of $\theta=0^{\circ}$ corresponds to the spacecraft viewing the flare at the disk centre while $\theta=90^{\circ}$ corresponds to viewing the flare at the solar limb. Here, HXR directivity is defined as the energy-dependent ratio,

\begin{equation}
    \text{HXR\;directivity}=\frac{\text{HXR\;flux\;at\;spacecraft\;1, \;small\;}\theta}{\text{HXR\;flux\;at\;spacecraft\;2,\;large\;}\theta}
\end{equation}
i.e., the ratio of X-ray energy spectra (see Figure \ref{fig_00}), where we define spacecraft 1 (SC1) as the spacecraft viewing the flare at a smaller heliocentric (observer) angle $\theta$ and spacecraft 2 (SC2) as the spacecraft viewing the flare at a larger angle $\theta$. We do this for consistency, the X-ray albedo component is always greatest near the disk centre (at small viewing angles) which leads to $>1$ directivity ratios at energies where the X-ray albedo is found, i.e., $\approx 20-90$~keV.

\begin{figure*}
    \centering
    \small{{\sffamily \textcolor{darkgray}{Dependence on Electron Directivity}}}\\
    \includegraphics[width=0.48\textwidth]{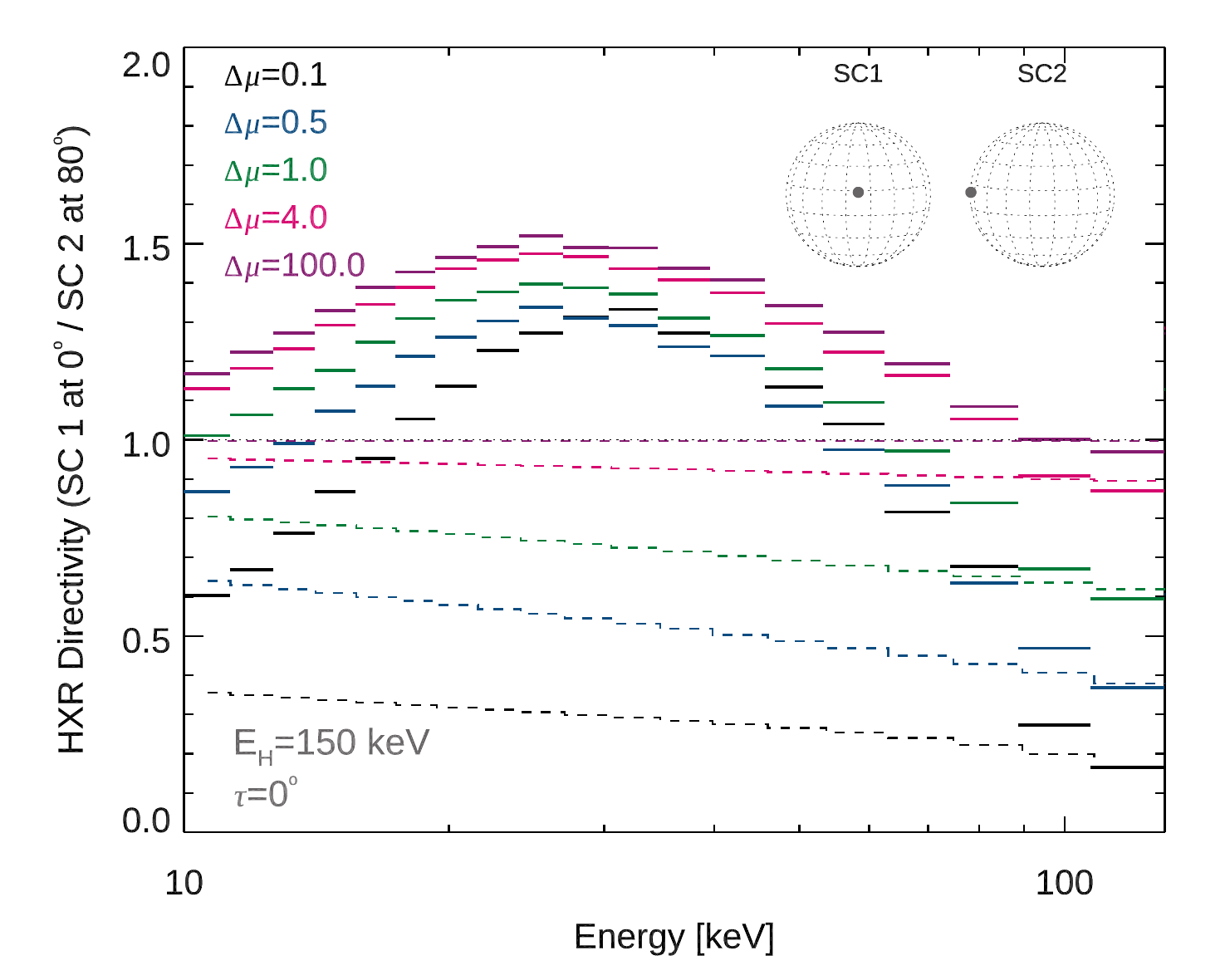}
    \includegraphics[width=0.48\textwidth]{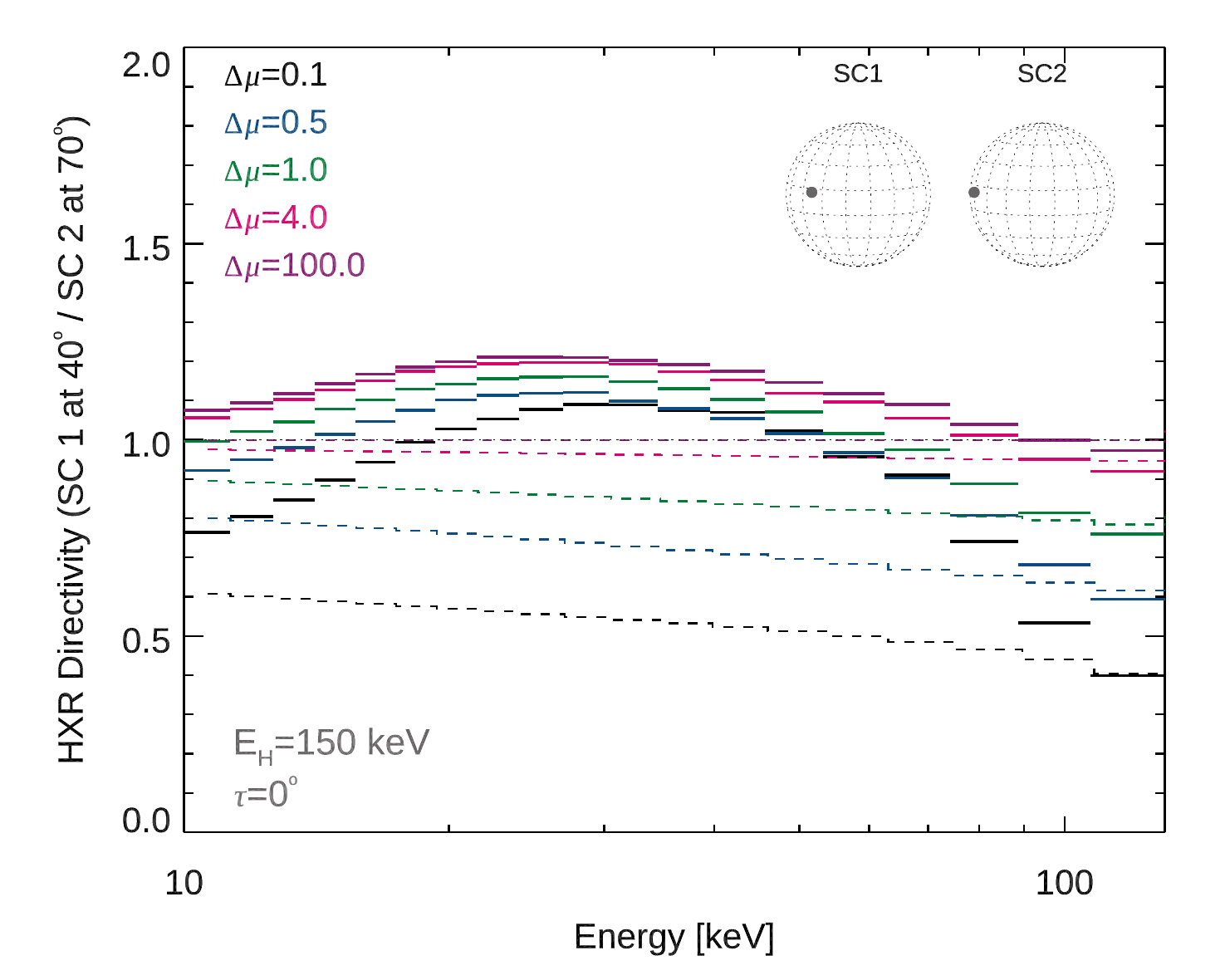}
        \caption{HXR directivity for spacecraft viewing angles of $0^{\circ}$ and $80^{\circ}$ (left) and $40^{\circ}$ and $70^{\circ}$ (right), showing how the HXR directivity changes with the directivity of the electron distribution (from very beamed $\Delta\mu=0.1$ to isotropic $\Delta\mu=100.0$). In these examples, we use an electron spectral index of $\delta_{FP}=2$ (footpoint), an electron high energy cutoff of $E_{H}=150$~keV, and a loop tilt of $\tau=0^{\circ}$. Solid bars: total observed flux ratio, dashed lines: primary component only, without albedo.}
        \label{fig3}
\end{figure*}

We also consider the local geometry via the loop tilt $\tau$ (\textbf in Figure \ref{fig_1}, panel b)). Following \citet{2008ApJ...674..570E}, the loop tilt is defined as the angle between the line connecting the loop apex and the local solar radial direction. When a closed flare loop is viewed side-on, $\tau=0^{\circ}$ if the loop apex sits directly along the local solar radial direction. In \citet{2008ApJ...674..570E}, it was shown that loops with tilt $\tau>0^{\circ}$ produce a polarization angle $\Psi>0^{\circ}$, and here we examine how $\tau$ affects the HXR directivity from stereoscopic observations. $\tau$ is an important parameter as it can help us to constrain the local magnetic geometry of the flare (or at least the average, dominant geometry in a flare with a complicated magnetic geometry).

\section{Selection of Modelling Results}\label{results}

All the results here are shown for spatially-integrated X-ray emission. Although STIX has imaging capabilities, it is likely that we will not have spatially resolved spectra from the second spacecraft viewing the flare (e.g., Fermi/GBM or PADRE).

\subsection{Effects of HXR albedo and chosen parameters of study}

\begin{figure*}[t!]
\centering
\small{{\sffamily \textcolor{darkgray}{Dependence on Electron High Energy Cutoff}}}\\
\includegraphics[width=0.48\textwidth]{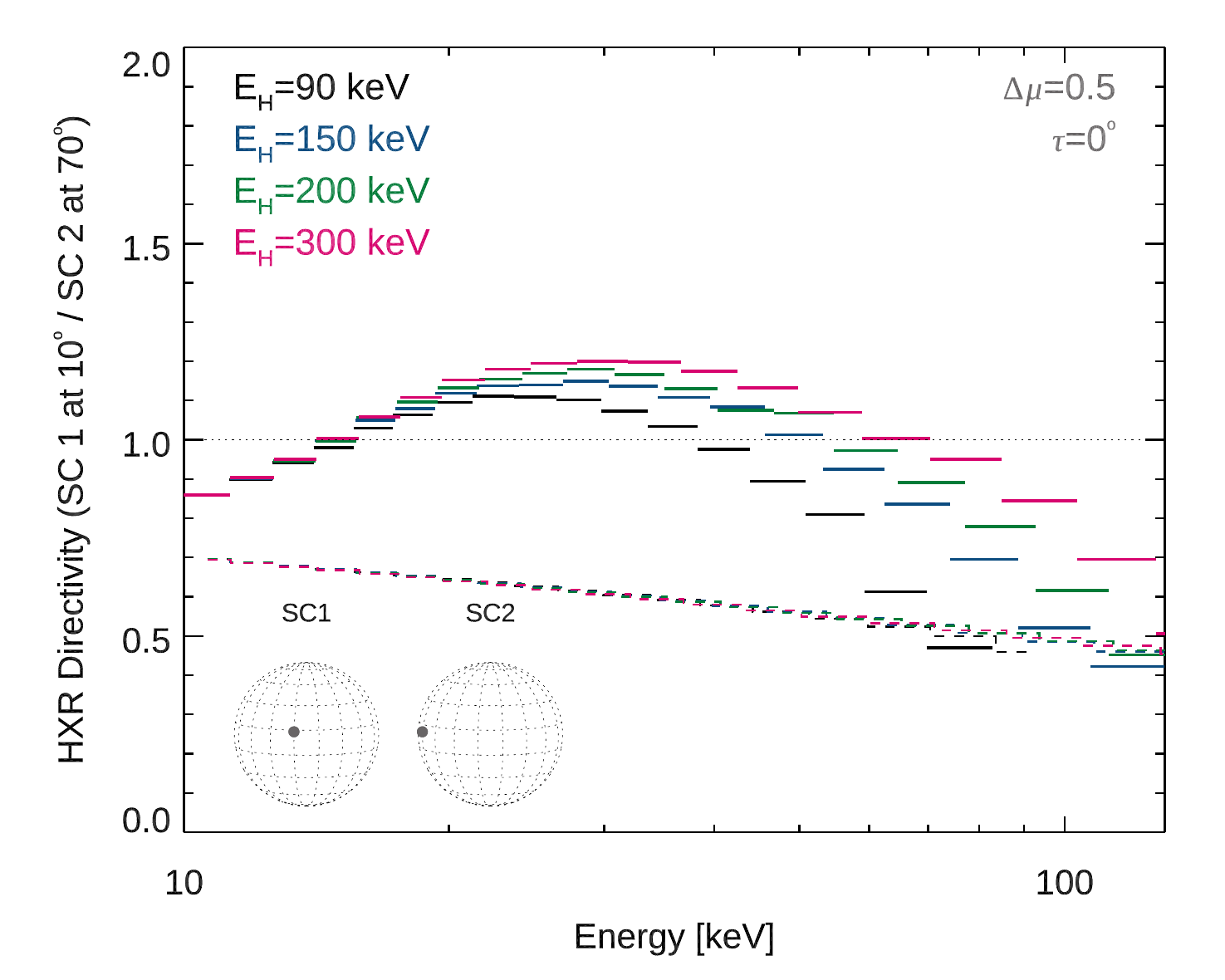}
\includegraphics[width=0.48\textwidth]{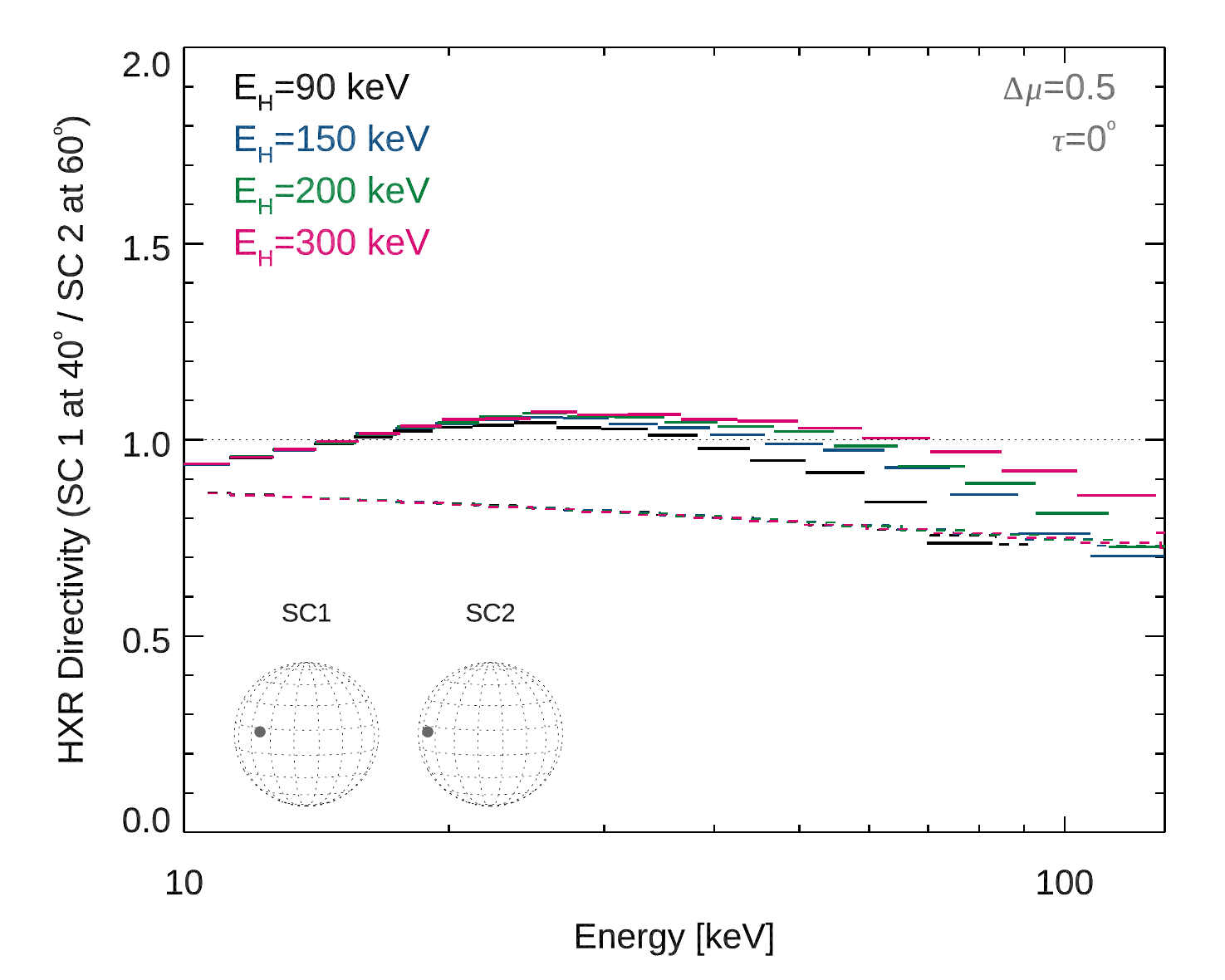}
\caption{HXR directivity for spacecraft viewing angles of $10^{\circ}$ and $70^{\circ}$ (left) and $40^{\circ}$ and $60^{\circ}$ (right), showing how the HXR directivity changes with the electron high energy cutoff (from $E_{H}=90$~keV to $E_{H}=300$~keV). In these examples, we use an electron spectral index of $\delta_{FP}=2$ (footpoint), an electron directivity of $\Delta\mu=0.5$, and a loop tilt of $\tau=0^{\circ}$. Solid bars: total observed flux ratio, dashed lines: primary component only, without albedo.}
\label{fig4}
\end{figure*}

\begin{figure*}[t!]
    \centering
    \small{{\sffamily \textcolor{darkgray}{Dependence on Loop Tilt}}}\\
       \includegraphics[width=0.48\textwidth]{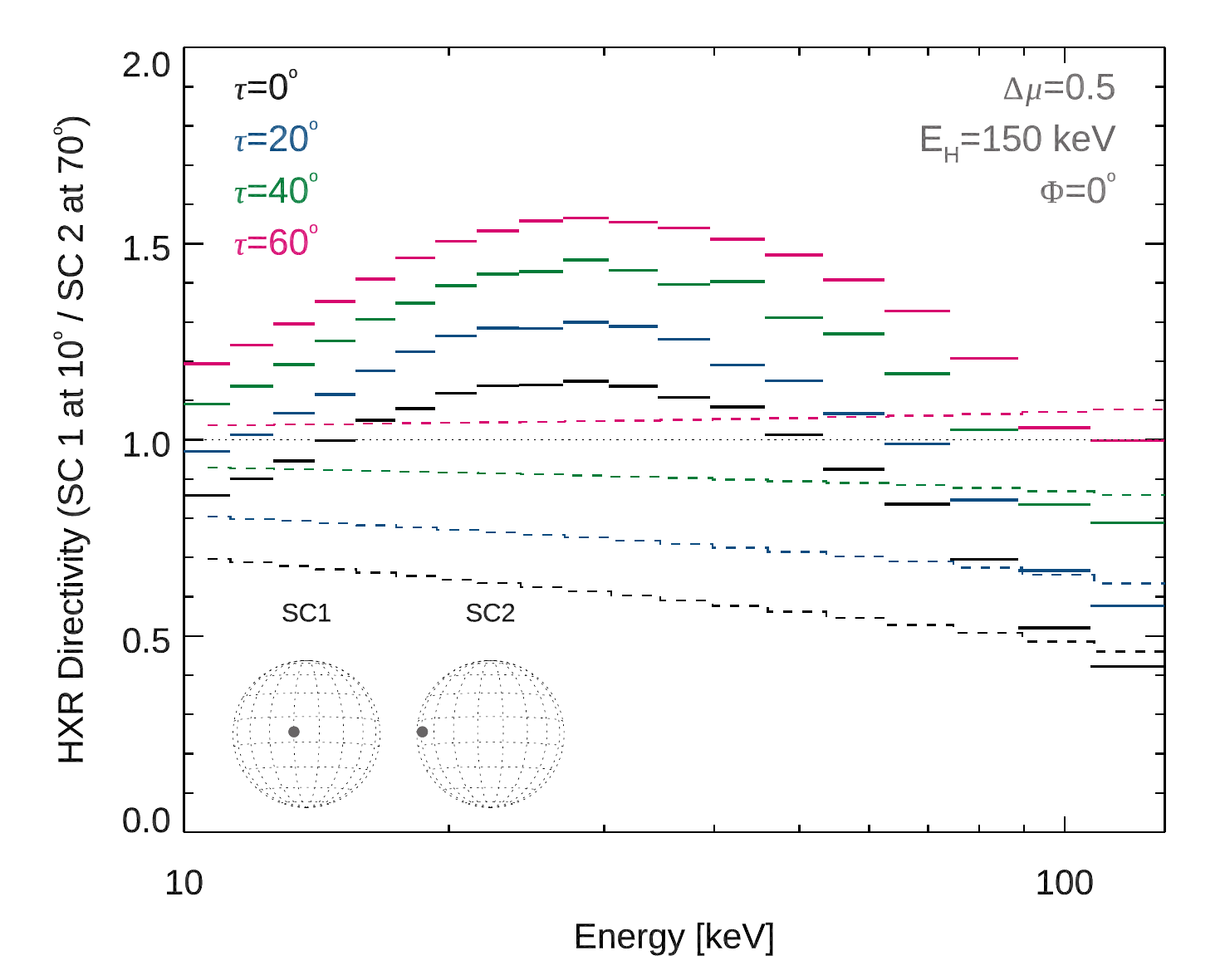}
    \includegraphics[width=0.48\textwidth]{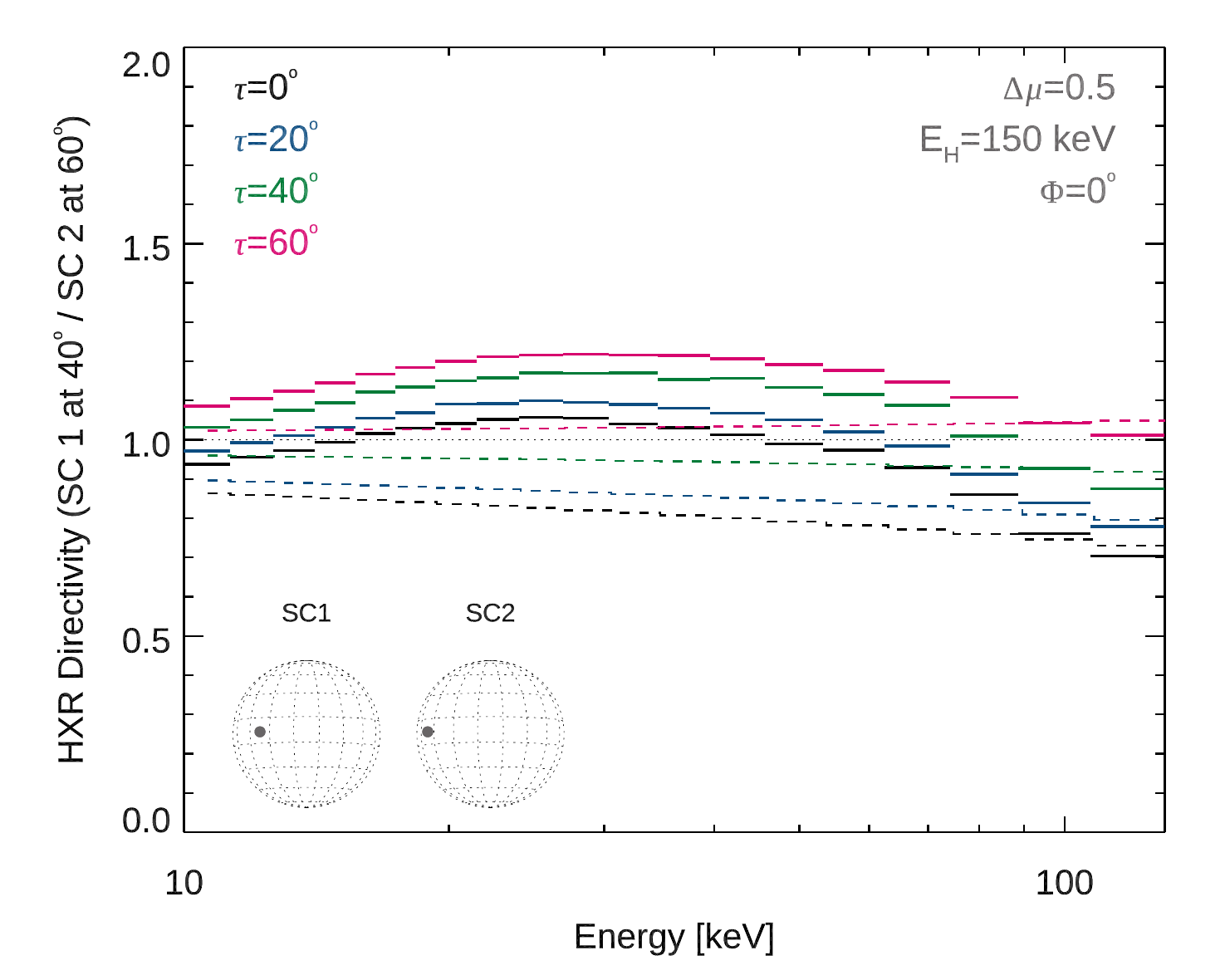}
        \small{{\sffamily \textcolor{darkgray}{Dependence on $\Phi$ for $\tau>0^{\circ}$}}}\\
           \includegraphics[width=0.48\textwidth]{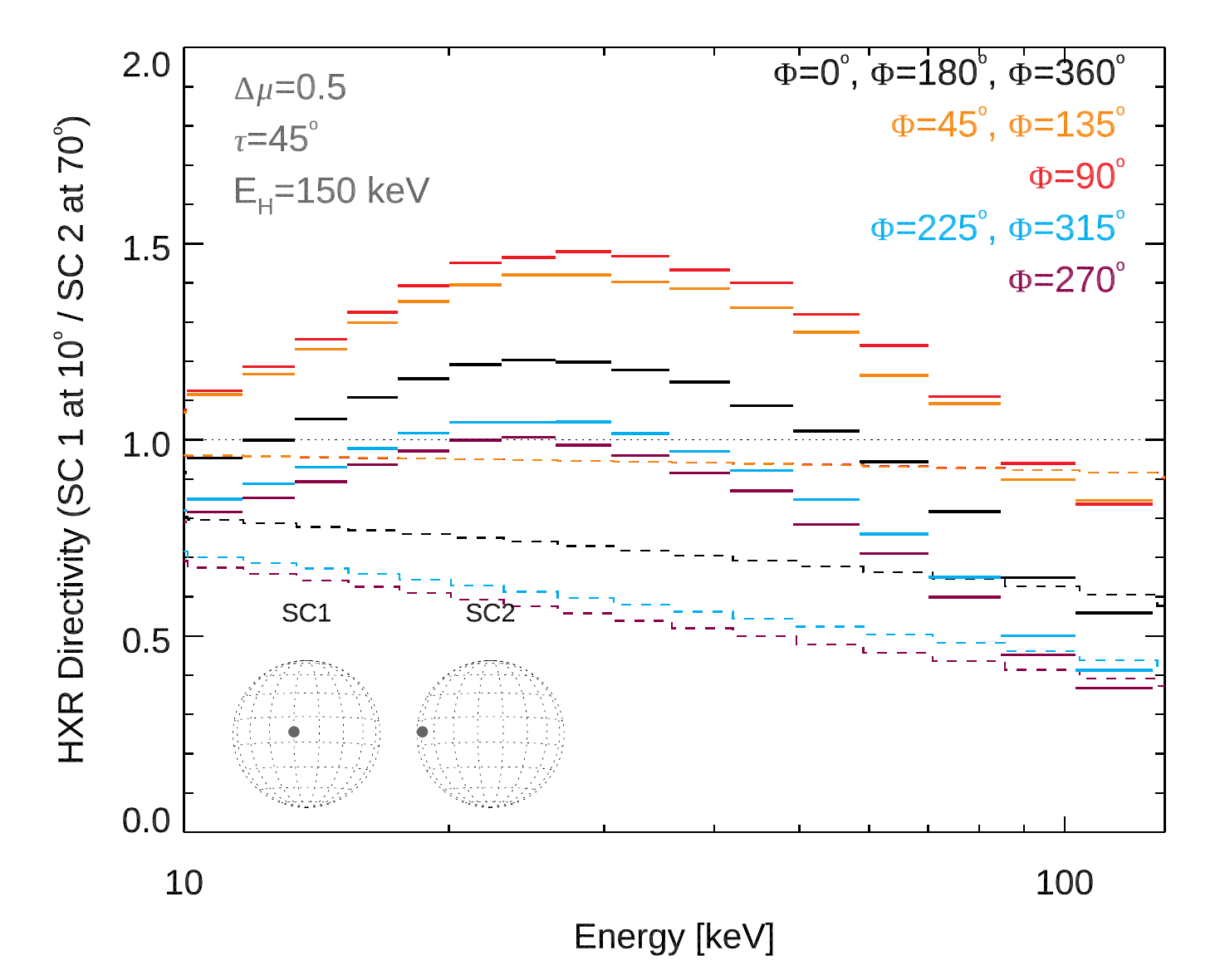}
    \includegraphics[width=0.48\textwidth]{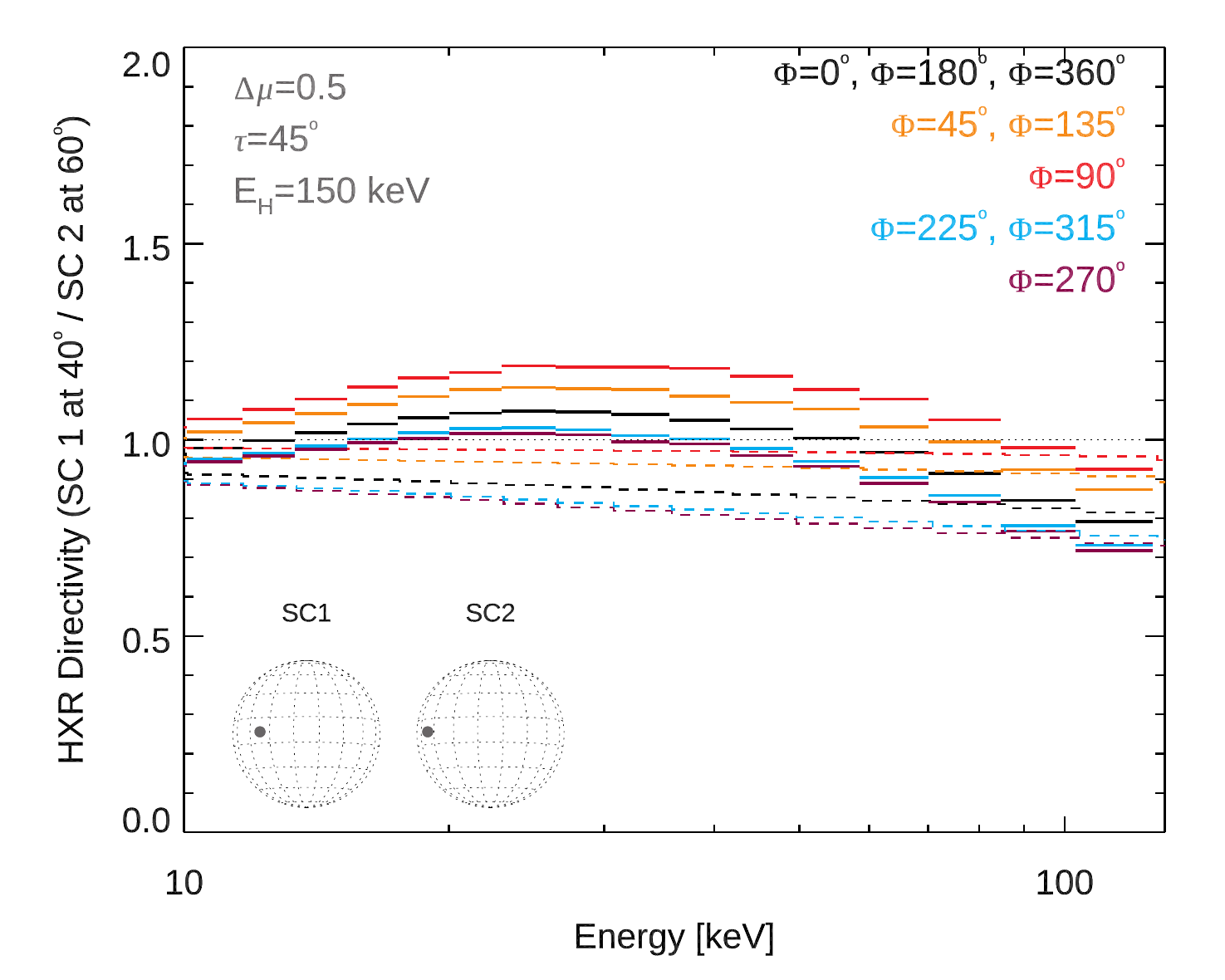}
    \caption{Top row: HXR directivity for spacecraft viewing angles of $10^{\circ}$ and $70^{\circ}$ (left) and $40^{\circ}$ and $60^{\circ}$ (right), showing how the HXR directivity changes with tilt of the magnetic loop away from the local solar radial direction (from $\tau=0^{\circ}$ to $\tau=60^{\circ}$). In these examples, we use an electron spectral index of $\delta_{FP}=2$ (footpoint), an electron directivity of $\Delta\mu=0.5$, and an electron high energy cutoff of  $E_{H}=150$~keV. Solid bars: total observed flux ratio, dashed lines: primary component only, without albedo. Bottom row: An example of how the HXR directivity can change at values of $\tau>0^{\circ}$ (here $\tau=45^{\circ}$) with $\Phi$, the azimuthal direction of the axis around which the loop is tilted \citep{2008ApJ...674..570E}.}
    \label{fig5}
\end{figure*}

Figures \ref{fig_00}-\ref{fig5} all show the effects of the X-ray albedo component and why it is so essential to account for this X-ray component before determining the HXR directivity and extracting electron and flare properties.

In general, the albedo component acts to increase the flux in small observer viewing angles, changing the HXR directivity ratio at energies between 10--100 keV from $<1$ (no albedo and assuming downward electron beaming) to $>1$ (with albedo). As already described, albedo is largest at smaller heliocentric angles. Flares with greater sunward directivity will produce a larger albedo fraction compared to isotropic distributions (e.g., \citealt{2011A&A...536A..93J}).

Flatter electron energy spectra (i.e., smaller spectral indices $\delta_{ FP}$) lead to higher albedo percentages and hence, larger HXR directivity ratios over the $\sim$20-70 keV range (Figure \ref{fig2}). The spectral index is a prime example of a non-thermal electron parameter that can be routinely extracted from single viewing angle spectral data, and hence, constrained before comparison with simulation.

Therefore, alongside the determination of electron directivity, we concentrate on examining parameters not easily constrained by current (single viewing angle) observations such as the:
\begin{itemize}
\item[-] Electron high energy cutoff $E_{H}$ which is related to the highest energy accelerated electrons and hence, also the properties of the acceleration mechanism and location.
\item[-] Loop tilt $\tau$ related to the local magnetic geometry.
\end{itemize}

In the majority of study cases, unless different properties are explicitly stated for that case, the following properties are used: $\Delta\mu=0.5$ (see Figure \ref{Fig0}), $E_{H}=150$~keV, $\tau=0^{\circ}$ and $\delta_{FP}=2$ (footpoint).

\subsection{Electron directivity}

Here we test five different electron footpoint directivities (see Figure \ref{fig3}) going from completely isotropic ($\Delta \mu = 100$) to very beamed ($\Delta \mu = 0.1$). Figure \ref{fig3} shows the resulting HXR directivities for spacecraft viewing angles of (left) SC1 at $0^{\circ}$  and SC2 at $80^{\circ}$, and (right) SC1 at $40^{\circ}$ and SC2 at $70^{\circ}$. While isotropic distributions produce the highest values of HXR directivity (due to albedo) over the $20-50$~keV range, very beamed distributions show the greatest HXR directivity change over the energy range of $10-100$~keV (comparing the dashed and solid lines), due to the larger albedo fraction produced by a large sunward directivity. At the curve peak (over the $20-50$~keV range), the difference between each directivity curve (beamed to isotropic) is at a minimum. 
At higher energies above 50~keV, the albedo dominance starts to diminish and the HXR directivity curve tends back to what we expect to see without the effects of albedo (dashed lines). Such trends stress the importance of using the entire observed X-ray energy range for the determination of electron directivity.
A comparison of the left and right panels in Figure \ref{fig3} demonstrates how the HXR directivity curves change for different spacecraft viewing angles. In general, larger differences in spacecraft viewing angles are preferable since they show the greatest difference between different electron directivites over all energies. However, Figure \ref{fig3} (right) demonstrates that even small differences in viewing angles (in this case $40^{\circ}$ and $70^{\circ}$) can still be used to extract the electron directivity, particularly if the flux uncertainties are small (large flare) and the whole energy range is used (this is discussed in subsection \ref{errors}).

In Figure \ref{Fig6}, we investigate the pancake distribution, where the peak electron flux is directed at $90^{\circ}$ to the guiding magnetic field. Interestingly, a directed pancake distribution (using $\Delta\mu=0.5$, see Figure \ref{Fig0}) can produce a HXR directivity spectrum comparable to a completely isotropic distribution, differing at the highest energies ($\ge 80$~keV) only.

\subsection{High energy cutoff}

As discussed, the HXR directivity curves are also sensitive to the `high-energy cutoff' $E_{H}$. Figure \ref{fig4} demonstrates how the HXR directivity changes for four different high-energy cutoff values of $E_{H}=90,150,200,300$~keV and each curve is different above 20~keV (in a real flare, it is likely that the isotropic thermal distribution will also dominate at $\le 20$~keV giving a directivity close to $1$). The greater the value of $E_{H}$, the flatter (or larger) the HXR directivity remains up until higher energies. When $E_{H}$ is lower, the lack of higher energy electrons creates an X-ray deficit in the $20-100$~keV energy range.

In Figure \ref{fig4}, we also compare the spacecraft observing angles of (left) SC1 at $10^{\circ}$ and SC2 at $70^{\circ}$, and (right) SC1 at $40^{\circ}$ and SC2 at $60^{\circ}$. Although the curves overlap at lower energies ($\le 30$~keV), individual curves are distinguishable even for the small differences in viewing angle ($20^{\circ}$, right). However for real data, such distinction will be highly dependent on the measurement uncertainties for the individual event.

\subsection{Loop tilt, footpoint orientation and magnetic field geometry}

In Figure \ref{fig5} (top row), we show an example of how the tilt of the flare loop (dominant magnetic field direction) can change the HXR directivity curves. For this example, we pick a high-energy cutoff value of 150~keV, a footpoint anisotropy of $\Delta\mu=0.5$ (mildly beamed) and spacecraft viewing angles of $10^{\circ}$ and $70^{\circ}$ (left) and $40^{\circ}$ and $60^{\circ}$ (right).

Figure \ref{fig5} (top row) shows the HXR directivity curves for four different loop tilts of $\tau=0^{\circ},20^{\circ},40^{\circ},60^{\circ}$. As the loop tilts, the direction of the X-ray emission with respect to the observer changes. From Figure \ref{fig5}, we can see that increasing the loop tilt $\tau$ has the effect of shifting the entire HXR directivity curve up to higher values.

The effect of the loop tilt is not only dependent on the heliocentric angle $\theta$ but on $\Phi$, the azimuthal direction of the axis around which the loop is tilted \citep{2008ApJ...674..570E} (azimuth of the footpoint line relative to the radial line; see Figure \ref{fig_1}, panel b)). In Figure \ref{fig5} (top row) and all other simulations shown in this paper, $\Phi=90^{\circ}$, which corresponds to a loop with footpoints oriented perpendicular to the direction from the disk center to the flare location.

$\Phi$ is similar to $\theta$, in the sense that it should be observable and hence, can be constrained using STIX and/or EUV imaging observations. Once $\theta$ and $\Phi$ are determined, directivity can be associated with the local magnetic loop configuration.

An example of how the HXR directivity can vary, using $\tau=45^{\circ}$, with different $\Phi$ ranging from $0^{\circ}$ to $0^{\circ}=360^{\circ}$ is shown in Figure \ref{fig5} (bottom row). Values of $\Phi<90^{\circ}$ and $\Phi>90^{\circ}$ produce an overall lower HXR directivity at all values of energy.

Determination of the loop tilt is useful, since properties regarding the magnetic configuration are often difficult to constrain from imaging alone. This result matches the work of \citet{2008ApJ...674..570E}, showing how the polarization angle (which will be measurable with PADRE) can also be used to determine the orientation of the loop.

In real observations, this property may be complicated by multiple loops tilted at different angles. Here, we are only interested in spatially integrated flare observations matching the current capabilities of (most) instruments, and hence the tilt parameter will indicate an overall average loop tilt configuration.

\subsection{Coronal transport effects; a brief discussion}

Transport effects such as collisions and turbulent scattering act to diminish directivity; collisions tend to remove directivity at lower energies, while turbulent scattering can isotropise and remove directivity efficiently at higher electron energies leading to completely isotropic distributions as shown for X-ray polarization in \citet{2020A&A...642A..79J}.

The inclusion of a background thermal component in flare X-ray spectra will also tend to hide any directivity signatures at lower energies below 20~keV since it is expected the thermal component will have a directivity close to isotropic. Other transport effects such as electron beam-induced return currents \citep[e.g., ][]{1977ApJ...218..306K,1980ApJ...235.1055E,2006ApJ...651..553Z,2017ApJ...851...78A,2021ApJ...917...74A} may also play a role but we do not attempt an exhaustive coronal transport study in this initial work.

As one example, we demonstrate the effects of coronal collisions using two simulation runs. The electron distribution is injected into a coronal loop apex as a power law with $\delta=4$ (corona) giving $\delta_{FP}\approx 2$ (footpoint). In the corona, we model collisional transport using a background temperature of $T=20$~MK, and a half loop length $L=20''$ (as seen from Earth, from the apex to chromosphere). Two values of coronal number density are used: $n=10^{9}$~cm$^{-3}$ and $n=10^{10}$~cm$^{-3}$.

\begin{figure*}[t!]
    \centering
    \includegraphics[width=0.49\textwidth]{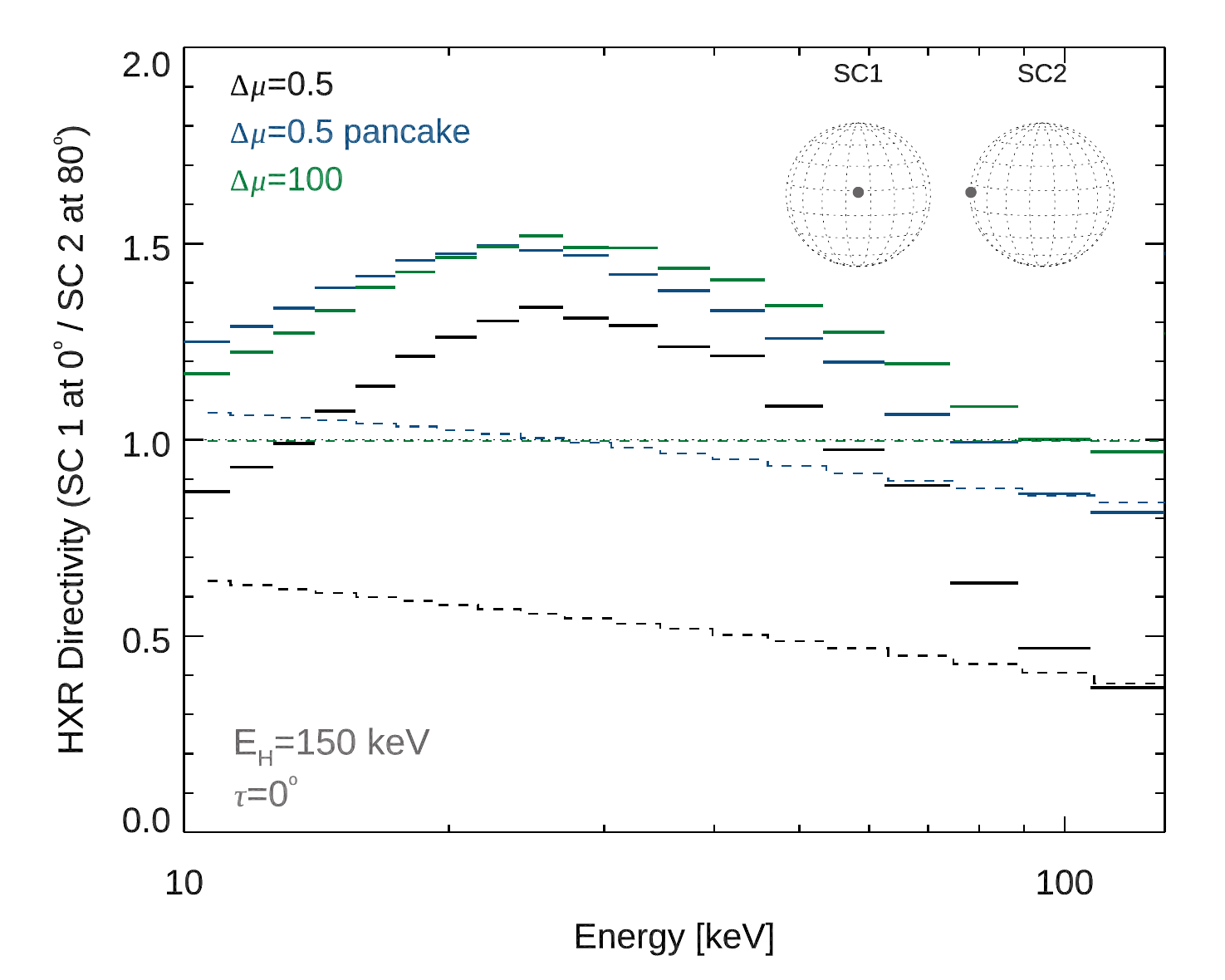}
    \centering
    \includegraphics[width=0.49\textwidth]{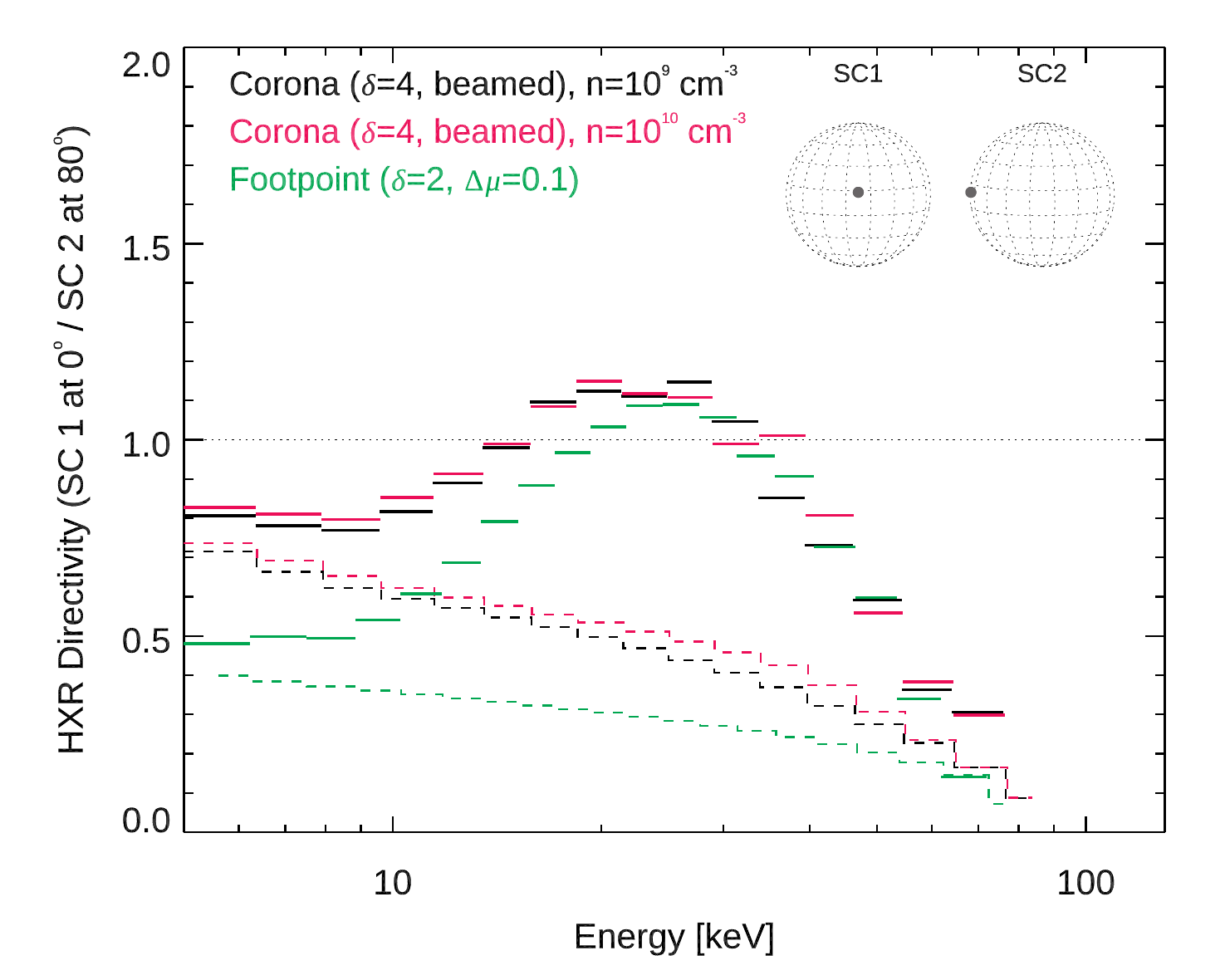}
      \caption{(Left) Comparison of the HXR directivity for a downward beamed ($\Delta\mu=0.5$) distribution, an isotropic distribution and a pancake distribution with the bulk of the electron velocity directed towards $90^{\circ}$. It is interesting to note that the HXR directivity curve from a pancake distribution can appear similar to an isotropic distribution. (Right) Effects of collisions in coronal plasma of different densities using densities of $n=10^{9}$~cm$^{-3}$ and $n=10^{10}$~cm$^{-3}$.}
      \label{Fig6}
\end{figure*}

The resulting HXR directivity curves are compared in Figure \ref{Fig6} (right). We compare the curves with a transport-dependent footpoint simulation using the following parameters: $\delta_{FP}=2$ (footpoint), $\Delta\mu=0.1$, $E_{H}=90$~keV and $\tau=0^{\circ}$. Here, we can clearly see that the transported electron distribution (black and pink curves) isotropise as expected, particularly at lower energies $<30$~keV, increasing the value of the primary (dashed lines) and total (solid bars) HXR directivity, compared to the resulting (similar) HXR directivity created by the electron distribution injected at the footpoint (green).

\begin{figure*}[hbtp!]
    \centering
    \includegraphics[width=0.49\textwidth]{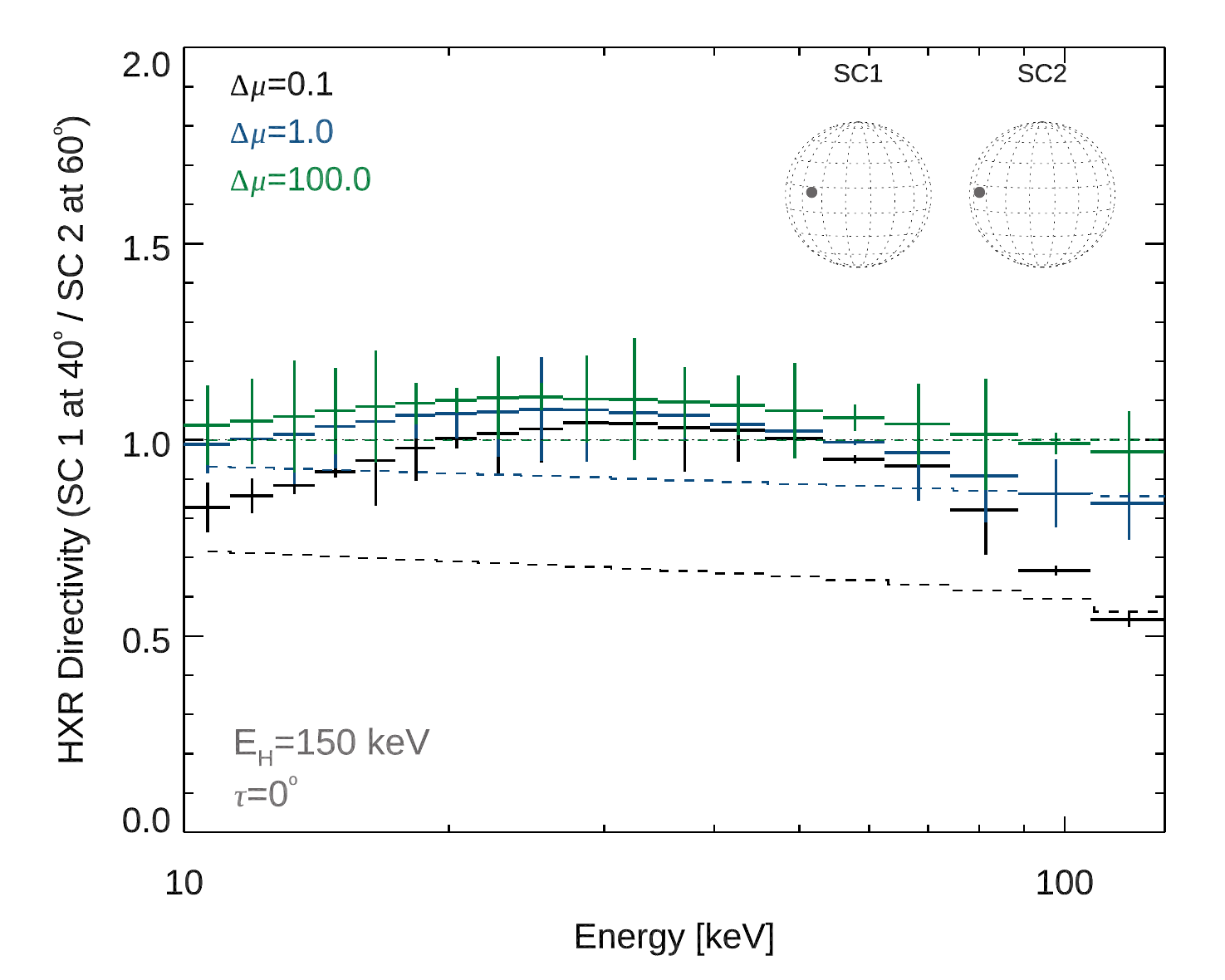}
      \includegraphics[width=0.49\textwidth]{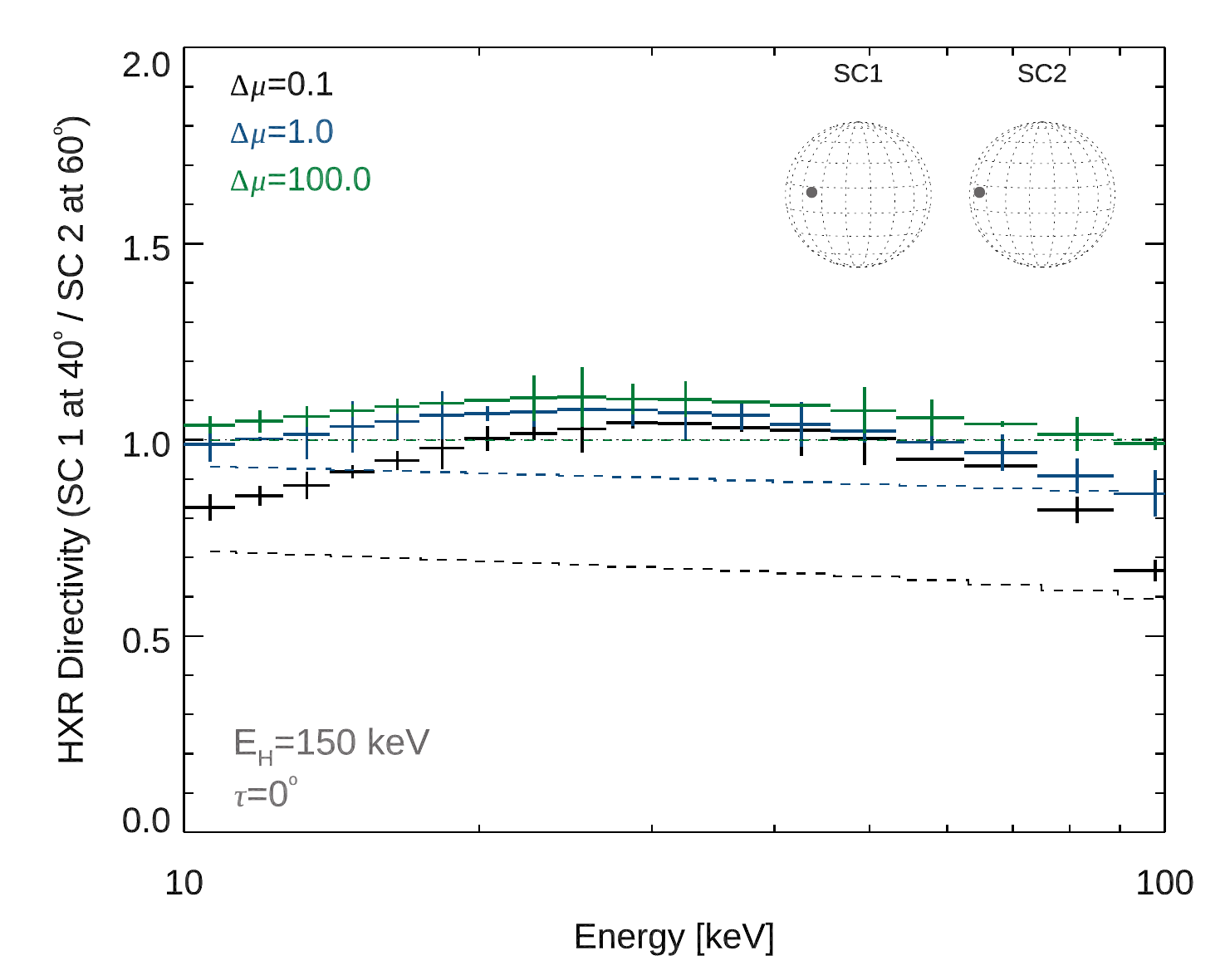}
    \centering
    \includegraphics[width=0.49\textwidth]{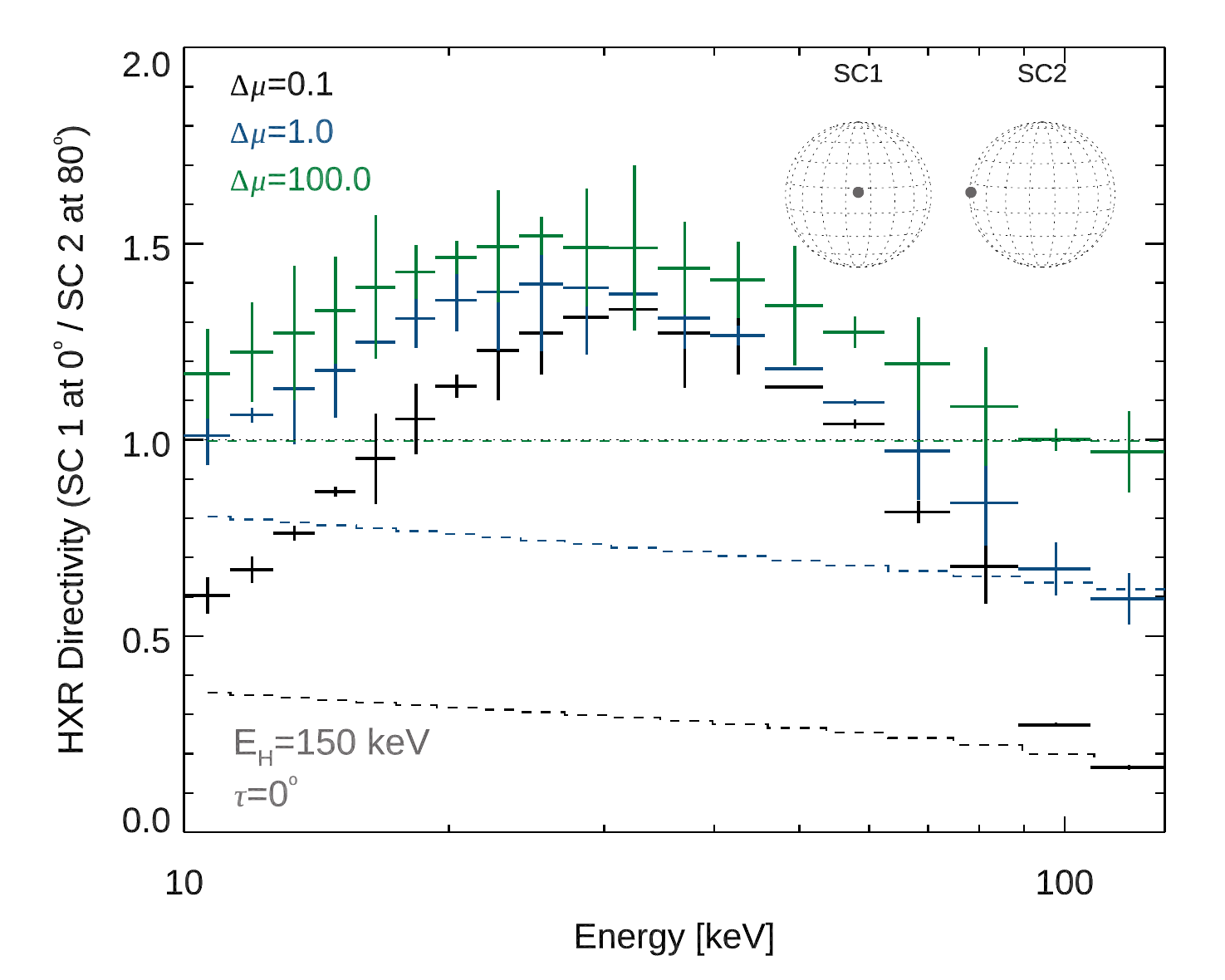}
      \includegraphics[width=0.49\textwidth]{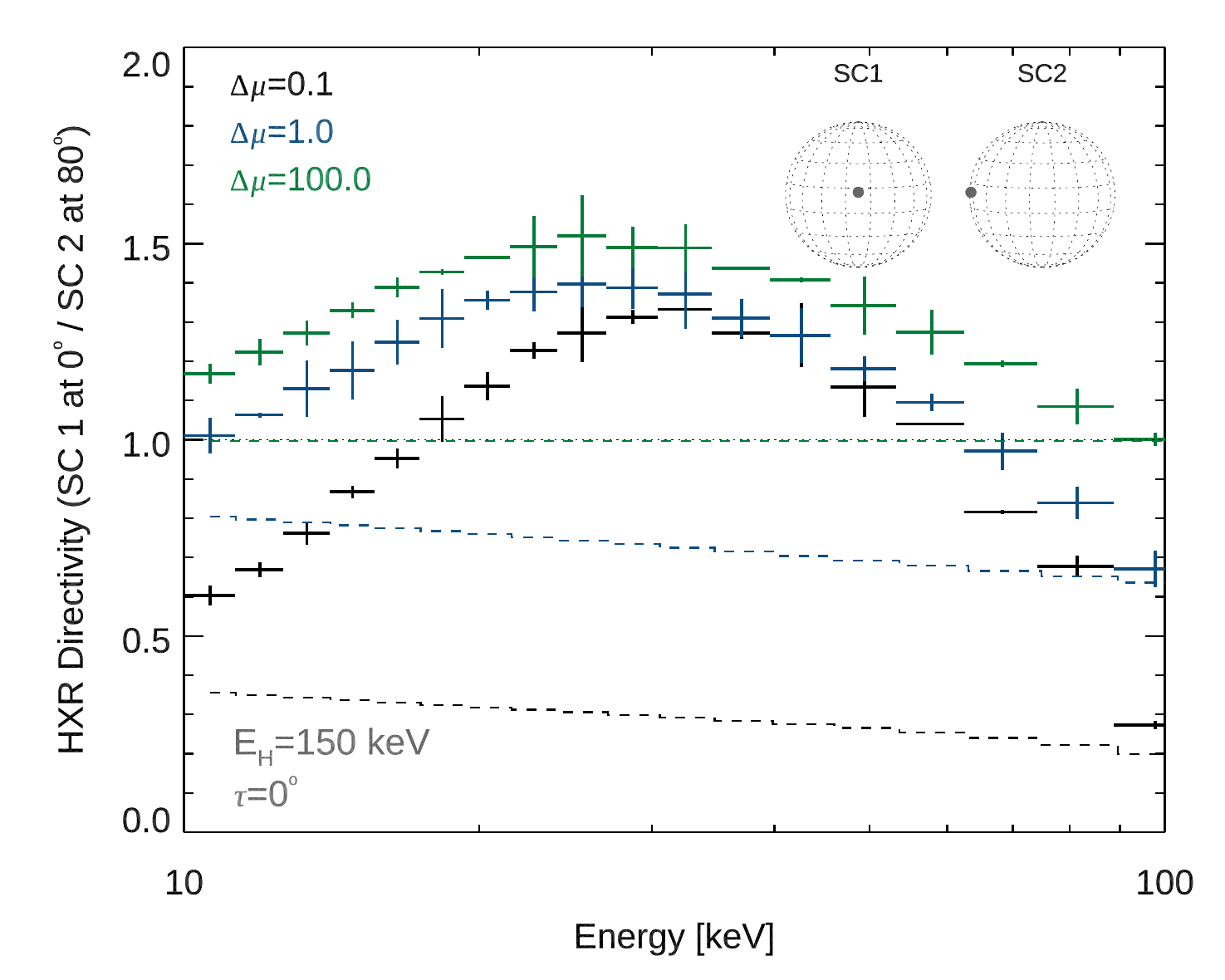}
      \caption{Top row: HXR directivity curves for small difference viewing angles of $40^{\circ}$ and $60^{\circ}$ created from individual simulated spectra using random uncertainties with a maximum of a) 10\% and b) 5\% in all energy bins. Bottom row: same as top but for large difference viewing angles of $0^{\circ}$ and $80^{\circ}$.}
      \label{uncer}
\end{figure*}

\begin{figure*}[hbpt!]
\centering
\includegraphics[width=0.99\textwidth,trim=25 0 70 00,clip]{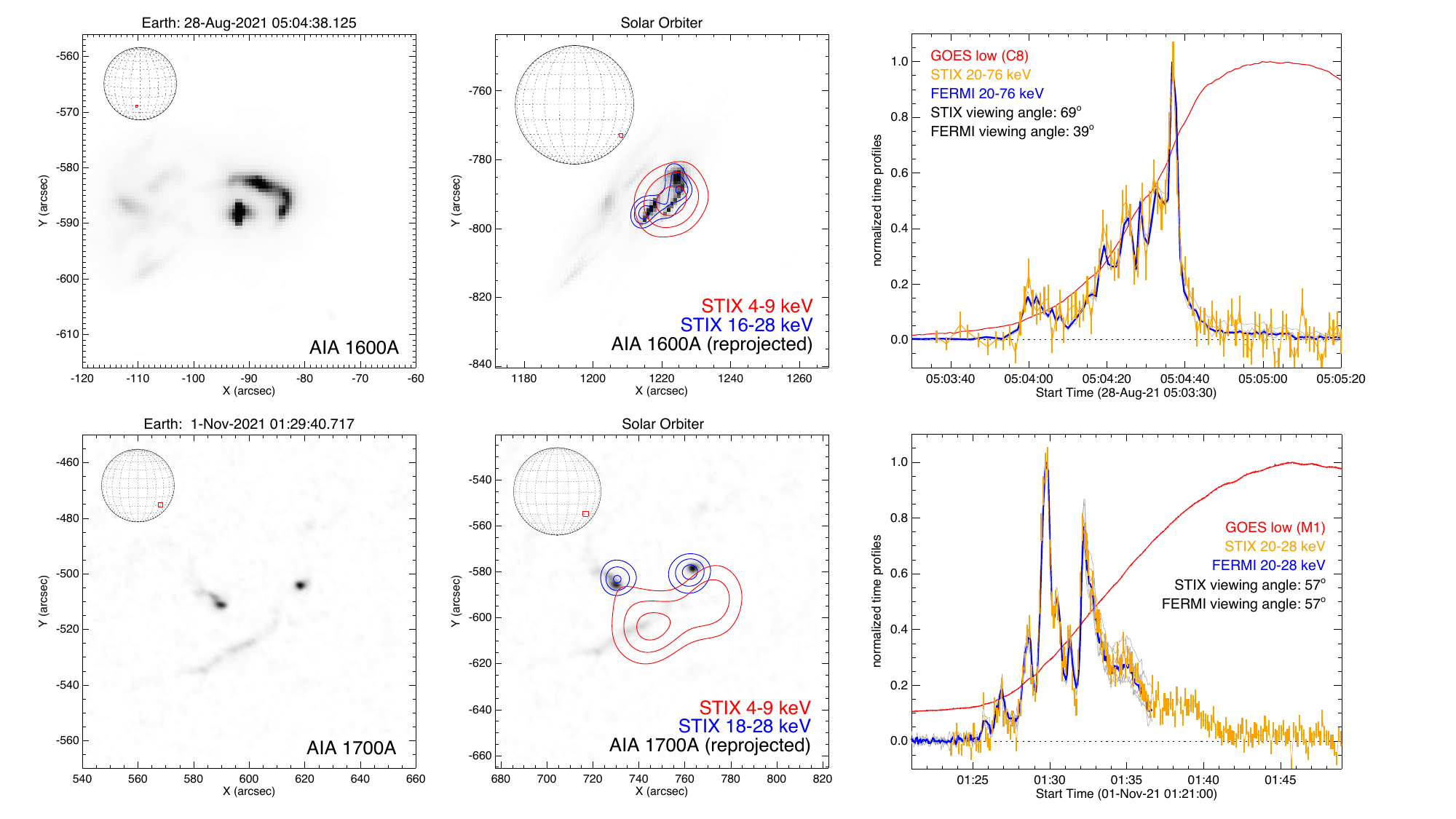}
\caption{X-ray images and time profiles of the two selected flares: (Top) flare SOL2021-08-28T05 where STIX and Fermi/GBM have clearly different viewing angles. (Bottom) flare SOL2021-11-011T01 is seen from essentially the same viewing angle for both instruments. The images show the Solar Dynamics Observatory Atmospheric Imaging Assembly (SDO/AIA) \citep{2012SoPh..275...17L} UV source configuration as seen observed from Earth (left), as well as reprojected to the SolO vantage points (center). Both flares show a good spatial match between the chromospheric UV emission and the non-thermal HXR signal. On the right, time profiles are shown. All curves are normalized to their maximum and therefore the directivity signal is not obvious in these overview figures. The STIX profiles have been shifted to account for the different light travel time to SolO. The individual Fermi/GBM detectors are shown in light grey and the detector-summed curve is shown in blue. The individual Fermi/GBM detectors give similar profiles, expect for the last part of the November 1 flare (SOL2021-11-011T01). This could possibly be an effect of pileup. For STIX, the profiles have been produced by summing over 24 of the 30 imaging detectors excluding the finest subcollimators. The time evolution of the non-thermal HXR component observed by the two instruments agree well within the error bars of STIX.}
\label{fig_obs}
\end{figure*}

\subsection{Effect of uncertainties}\label{errors}

In order to account for the uncertainties associated with real X-ray data and the limits of HXR directivity determination, we examine two simple examples in Figure \ref{uncer} where the individual spectra are assigned random flux uncertainties with a maximum of either 5\% or 10\%.

For small differences in viewing angle ($\approx20^{\circ}$), at high heliocentric angles ($>40^{\circ}$), large uncertainties ($\approx 10\%$) would obscure the directivity but 5\% uncertainties could allow differences at low (10-20 keV) and higher energies ($>80$~keV) to show, allowing the observer to extract whether the distribution is isotropic or has some level of directivity at the very least.

For large differences in viewing angles, 5\% uncertainties should allow the directivity to be constrained to a greater degree, while 10\% should will allow us to distinguish between `beamed' and `isotropic' electron anisotropies.

The errors in the observed X-ray fluxes have three main contributing factors: counting statistics, the accuracy of the pre-flare background subtraction, and systematic errors in the calibration of the instrument. In the low-energy range where the signal strength is high and well above the background level, systematic errors dominate the total errors in the flux. By comparing flare observations with similar viewing angles, the bias of systematic errors can be evaluated. In section \ref{prem}, we show that systematic errors between STIX and FERMI/GBM are around 6\% for our preliminary analysis of a single event. As the observed HXR spectra are steep, fluxes at higher energies are much weaker and eventually become of similar strength or even below the pre-flare background level. Hence, for high energies the error is given by a combination of counting statistics and/or the accuracy of the background subtraction. In practice, this leads to an upper limit in the photon energy for which errors stay below a scientifically meaning value.

\section{Preliminary comparison with HXR data}\label{prem}

While it appears to be straight-forward to determine X-ray directivity by measuring two absolute fluxes from two different vantage points, the difficulty of establishing an accurate absolute flux calibration in X-rays makes directivity measurement extremely challenging. To illustrate the difficulty of absolute X-ray calibration, a published comparison of X-ray instrumentation can be consulted. \citet{2005SPIE.5898...22K} compared Crab nebula observations by many different astrophysical X-ray observatories and found that deviations of 10$\%$ are frequently seen. For solar observations, the same difficulties remain, although we have two strong methods to check the validity of the absolute flux calibration: (1) thermal X-ray emission which is typically seen below 20 keV is expected to be isotropic, at least within our targeted goal of a few percent accuracy in the absolute flux measurement. Hence, the measured directivity of the thermal emission should be unity within the achieved accuracy, and (2) as a control group, we can use flares for which the viewing angles are similar for two spacecraft. Each time SolO is crossing the Earth-Sun line, we will get a few days where the look direction is similar to Earth-orbiting observatories. Alternatively, we can select flares which occur at angles in-between the two spacecraft. As the Fermi/GBM calibration is less accurate in the thermal energy range compared to the non-thermal range, here we will use the second approach. Despite having these consistency checks, the initial step is to do the best possible calibration with an accurate measurement of the associated errors. 

\begin{figure*}[hbpt!]
    \centering
\includegraphics[width=0.99\textwidth,trim=20 0 250 00,clip]{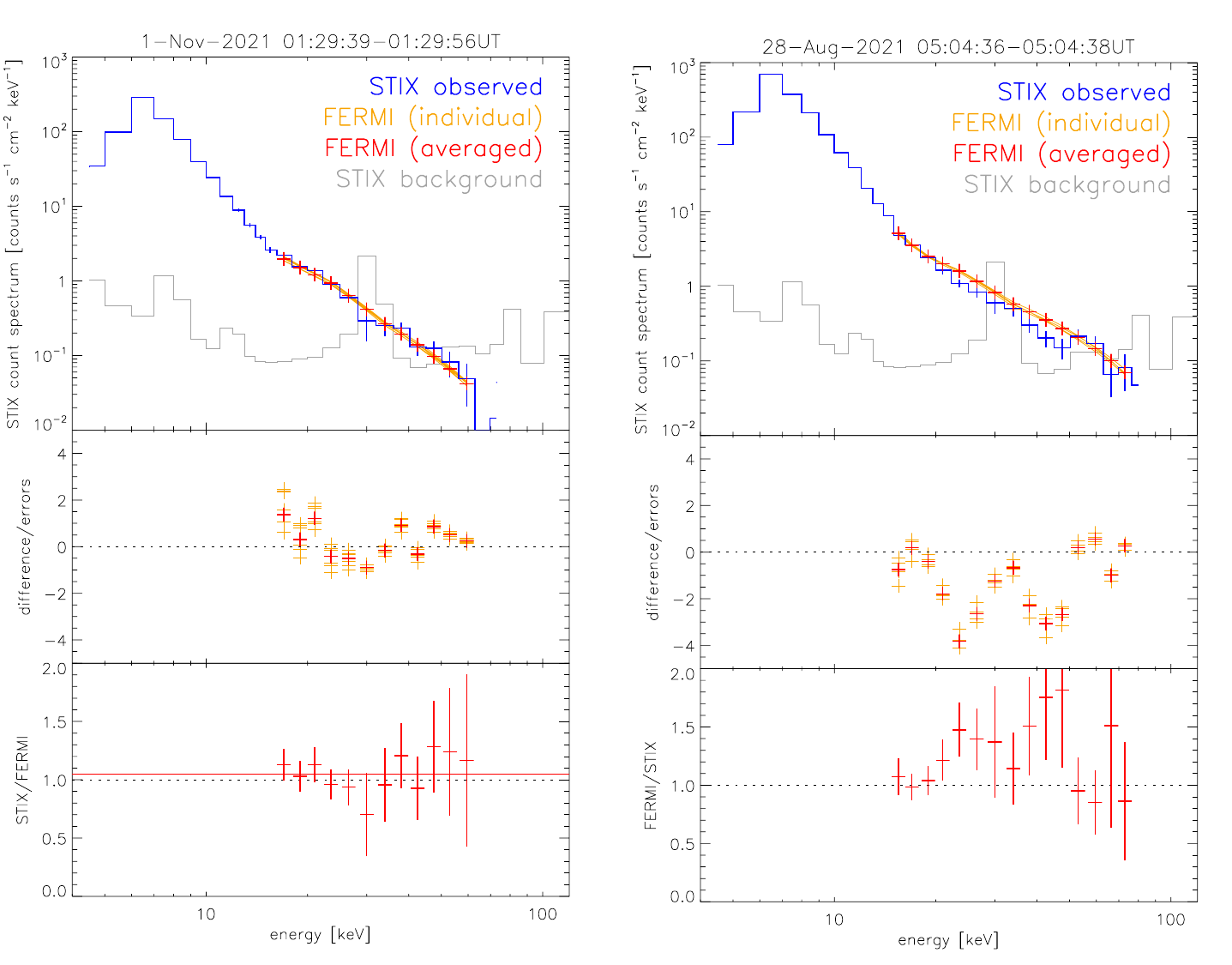}
    \caption{Comparison of the observed HXR directivity determined from Fermi/GBM and STIX observations for two flares: The flare on the left is seen from essentially 
    the same viewing angle (SOL2021-11-011T01), while the flare on the right has a difference in viewing angle of $\sim30^{\circ}$(SOL2021-08-28T05). For SOL2021-11-011T01, we find a directivity close to unity (1.05$\pm$0.06), as expected, validating the STIX and Fermi/GBM instrumental calibration. While the individual error bars are large, SOL2021-08-28T05 shows a systematic deviation from a directivity of unity, showing the expected albedo `bump'.}
    \label{direct_compare}
\end{figure*}

\begin{figure*}[hbpt!]
    \centering
\includegraphics[width=0.49\textwidth]{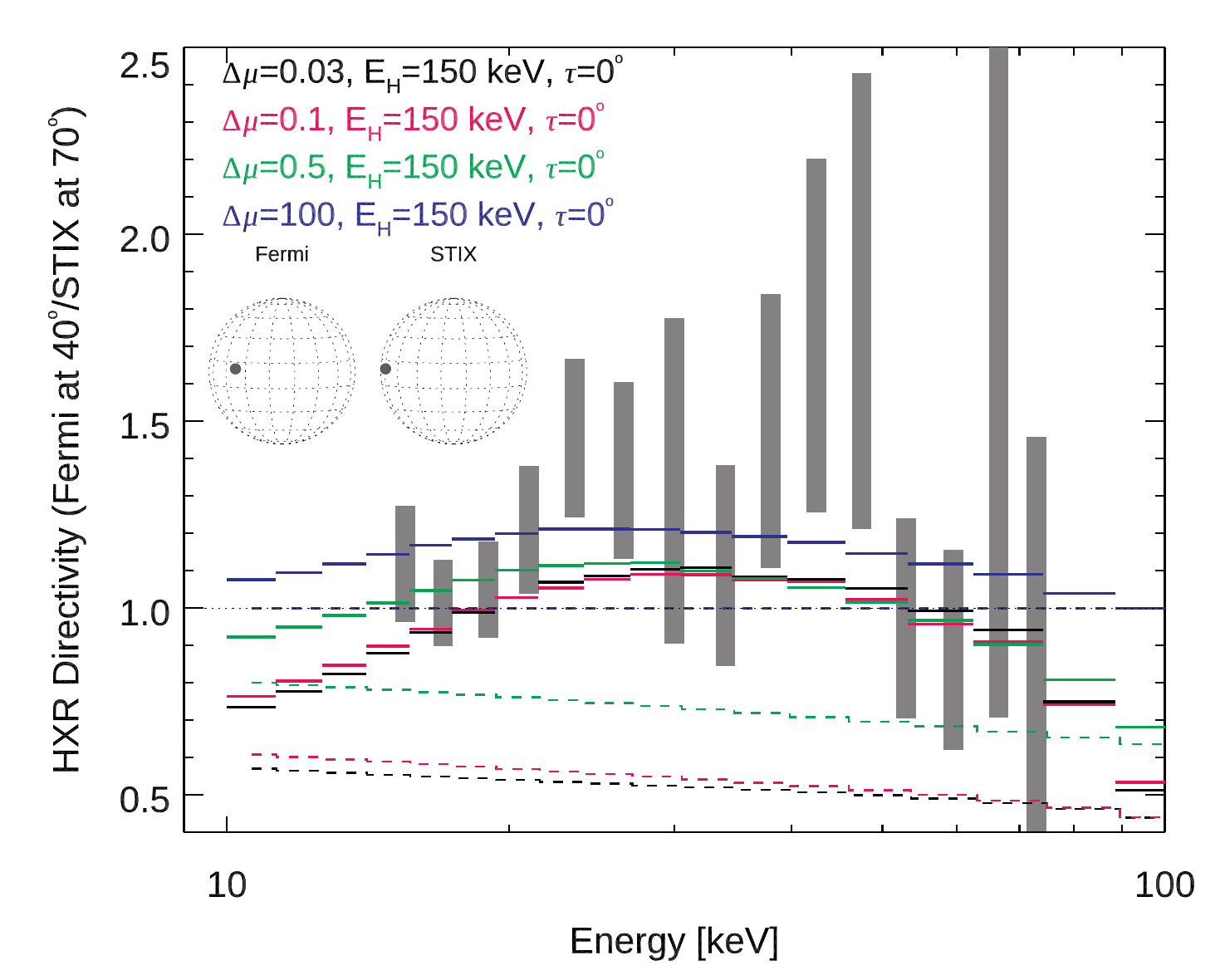}
\includegraphics[width=0.49\textwidth]{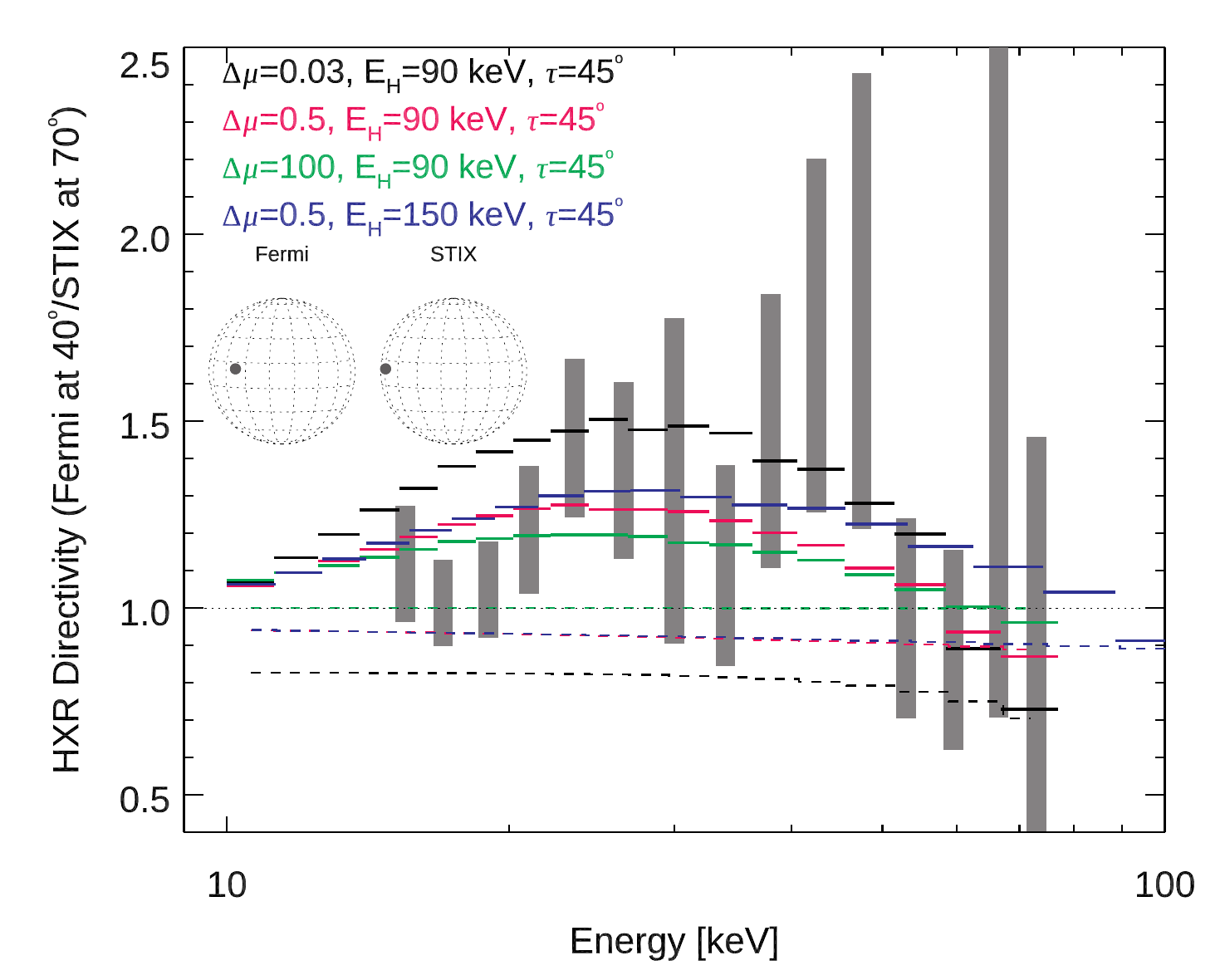}
\includegraphics[width=0.49\textwidth]{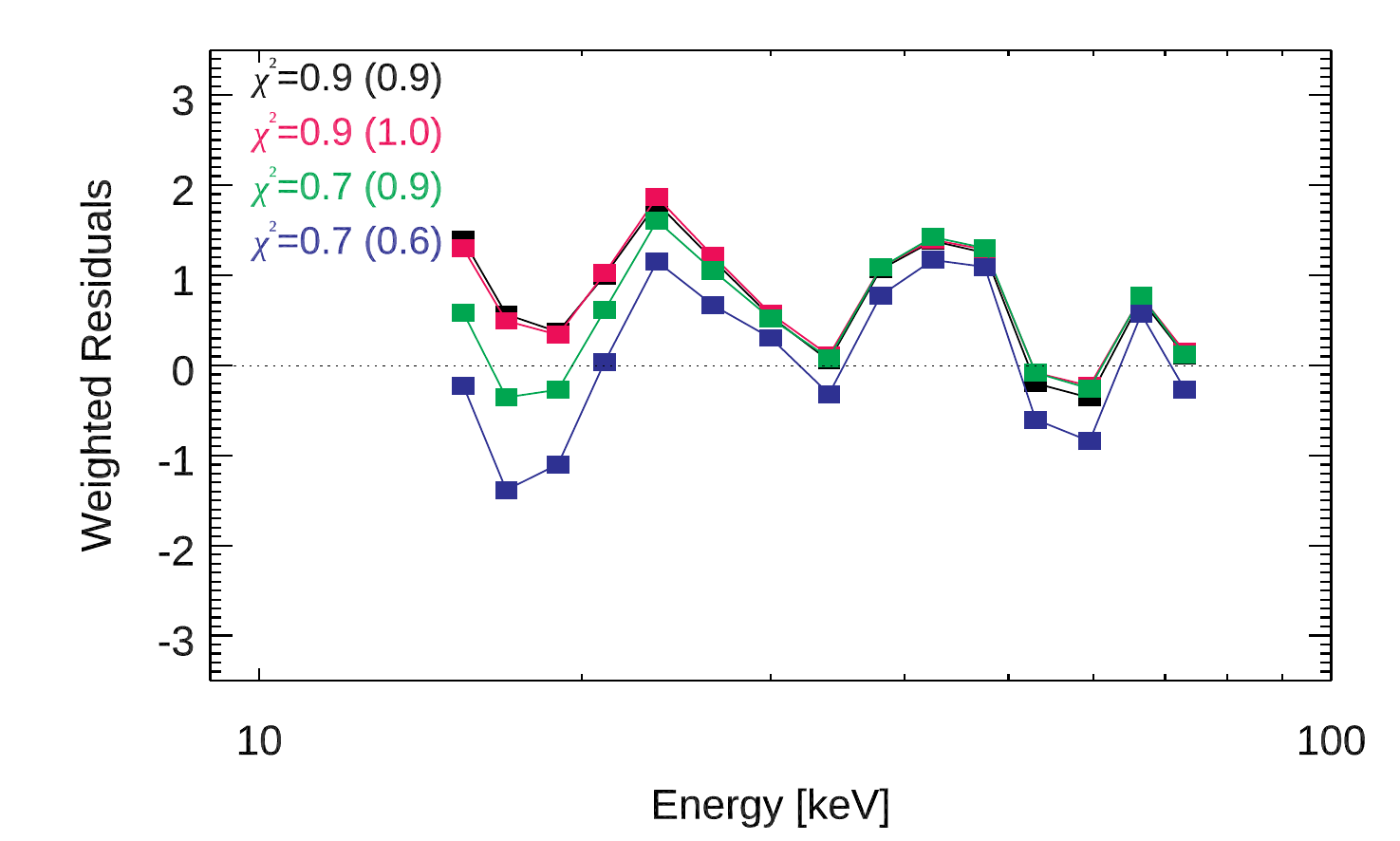}
\includegraphics[width=0.49\textwidth]{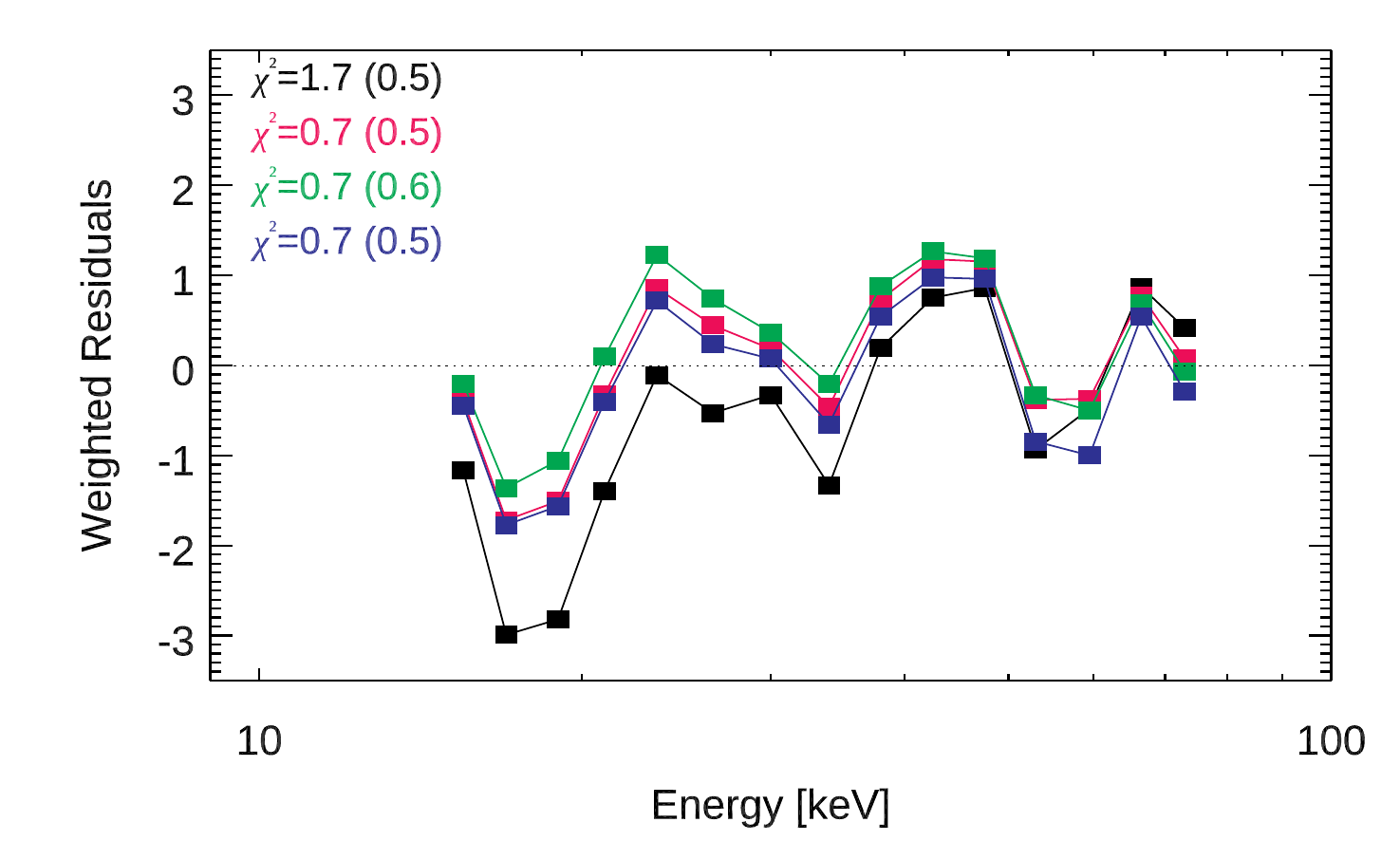}
    \caption{For flare SOL2021-08-28T05, a comparison of the HXR directivity determined from Fermi/GBM and STIX observations with selected simulation runs. Top left: determined HXR directivity (grey bars) and four simulation curves with varying electron directivities ($\Delta\mu = 0.03$, $\Delta\mu = 0.1$, $\Delta\mu = 0.5$ and $\Delta\mu = 100$) and $E_{H} = 150$ keV and $\tau = 0^{\circ}$. Top right: as top left but for simulation curves with $\tau = 45^{\circ}$ \textbf{($\Phi=90^{\circ}$)} and different electron directivities ($\Delta\mu = 0.03$, $\Delta\mu = 0.5$ and $\Delta\mu = 100$) and high energy cutoffs $E_{H} = 90$ keV and $E_{H} = 150$ keV. Bottom: The weighted residuals, i.e., (data-model)/(data error) for each simulation shown in the top row, alongside the reduced $\chi^{2}$ values for all energies and only energies above $23$~keV (value in the brackets).}
    \label{compare_1}
\end{figure*}

STIX has reached a satisfactory calibration level for most science topics. For directivity studies we must implement a few further corrections and then publish the results including their accuracy. This includes for example the `fuzziness' of the STIX grids as a function of photon energy. Currently the grids are assumed to have perfectly sharp edges, which is obviously a simplification. As the fine tuning of the STIX calibration is ongoing, here we show initial results with the currently available accuracy of the absolute flux calibration. These results give insight regarding what we observe within the currently available calibration, but they should not yet be used to draw stringent quantitative conclusions on electron beaming. What we present in this section should therefore be seen as an outlook to future directivity results. In Figure \ref{direct_compare} we show the HXR directivity curves for two flares, one for which the flare is seen from a very similar look direction (SOL2021-11-011T01), and a second one for which the viewing angle is separated by $\sim 30^{\circ}$ (SOL2021-08-28T05). As we only intend to show that such observations are feasible without drawing quantitative conclusions, we will not discuss the details of each flare in this paper. 

For our preliminary analysis of STIX and Fermi/GBM directivity measurements we selected two flares at moderate intensity to make sure that pileup effects are not an issue with Fermi/GBM. One of the flares (SOL2021-11-01T01; see Figure \ref{fig_obs}, bottom row) is seen from almost the same viewing angles for both instruments, and it is used to check that indeed a directivity of 1 is measured. The data is taken from a time when STIX was close to the Earth-Sun line. The viewing angle for both instruments is $57^{\circ}$. The second event (SOL2021-08-28T05; see Figure \ref{fig_obs}, top row) shows a $30^{\circ}$ difference in viewing angles. To simplify the initial analysis, the directivity is determined at the main non-thermal peak of each of the flares. We selected the following simple approach to measure directivity to obtain initial results:
\begin{itemize}
\item For each of the Fermi/GBM detectors, a time dependent background is subtracted by selecting pre- and post-flare time intervals. We then fit the time evolution using a polynomial. SOL2021-11-01 clearly has a time varying background and a polynomial of order 3 is used, while for SOL2021-08-28 the background is only minimally changing in time and a first order polynomial represents the background well. 
\item For each detector, the GBM spectrum is fitted over the peak time of the non-thermal emission using the Object
Spectral Executive (OSPEX software, \citealt{2002SoPh..210..165S}). To keep it simple, we only fit the non-thermal emission, excluding energies below 15 keV that contain thermal counts. The spectral form used in the fitting does not really matter for the initial analysis, as long as the fit represents the data well with even distributed residuals and a $\chi^2<2$. We use a thermal and a thick-target broken power law. Hence, in this initial analysis we do not yet fit an albedo component; we simply want to create a fit that represents the data well to be able to convolve the fitted function in the next step. 
\item The STIX spectral response matrix is used to convolve the GBM-derived fit parameters to calculate the expected counts in the STIX science energy bins. We use the standard deviation of the results of the individual GBM detectors as a measurement of the accuracy of the expected STIX counts. 
\item The ratio of the observed and the GBM-derived STIX counts is then calculated including the error estimates.
\end{itemize}

The main step that makes our analysis preliminary is the version of the STIX spectral response matrix used. The used version is from October 2023 and includes:
\begin{itemize}
    \item The Energy LookUp Table (ELUT, \citealt{2020A&A...642A..15K}) correction used assumes a constant spectral shape. The spectral shape is obviously not flat. However, the associated uncertainties are of the order of a percent for each individual pixel. As we average over hundreds of pixels, the error is smoothed out making the approximation of a flat spectrum viable. 
    \item The livetime correction uses parameters obtained by self-calibration to make the calibration line at 81 keV constant in time during flares. As the reported flares have rather high livetimes (i.e., above 97\% for the two events discussed), any further improvement of the livetime calculation will not affect the livetime correction used here. 
    \item STIX grids are assumed to have sharp edges (i.e., fuzzy grid implementation is not yet used). To minimize the effect, we only use the coarse half of the subcollimators. Estimates suggest that the introduced error is of the order of $\sim3$ percent at low energies (averaged over the coarse grids), but the error is not yet determined for the energy range relevant here. 
    \item The detector response matrix is at nominal STIX science bin resolution (i.e., without oversampling in energy). 
\end{itemize}
In any case, the currently used spectral response matrix should be well within the 10\% error limit that is desired for directivity measurements (see Section \ref{errors}). As it is difficult to precisely quantify the current shortcomings without implementing associated corrections, here we use our control flare which is seen from the same viewing angle for both instruments to showcase the currently available accuracy. The spectral analysis of our test case is shown in Figure \ref{direct_compare} (left) and there is indeed good agreement as expected: the average over all points give a ratio of 1.05$\pm0.06$. As the event is not particularly strong for STIX, the error bars are rather large above 30 keV. However, the measurements below 30 keV are well constrained. The individual Fermi/GBM detectors give the same result with a scatter of about 5\%. In summary, the current calibration has not yet reached the 5\% level for ratio measurements of individual energy bins, but the accuracy is better than 10\%.

For SOL2021-08-28T-05 (Figure \ref{direct_compare}, right), a directivity signal is measured. While the error bars at individual energies are rather large, it is clear that the Fermi/GBM detected signal is clearly stronger from $\sim$20 to $\sim$50 keV, as expected from the varying importance of albedo. For energies below 20 keV and above 60 keV, directivity is smaller. Below 20 keV, the measurements are getting consistent with unity. This could be due to the decreasing importance of albedo, but also the increasing contribution of the thermal X-ray flux, which is expected to be isotropic. The error bars at higher energies are very large, but nevertheless a trend is seen towards directivity $<1$ compared to what is seen at 30 keV. In the following paragraph, a more detailed comparison with the modeling results from the previous section is discussed. 

For SOL2021-08-28T05, we compared the observed HXR directivity with several transport-independent simulations using electron anisotropies of $\Delta\mu=0.03, 0.1, 0.5, 100$ (very beamed to isotropic), electron high-energy cutoffs of $E_{H}=90$~keV and $E_{H}=150$~keV, and a local loop tilt of either $\tau=0^{\circ}$ or $\tau=45^{\circ}$ (with $\Phi=90^{\circ}$). The residuals, calculated as (data-model)/(data error) and resulting reduced $\chi^{2}$ values, are then used to give an indication of ``goodness of fit''. For a selection of simulation runs, the data-simulation comparison, residuals and $\chi^{2}$ values are shown in Figure \ref{compare_1}. The results of Figure \ref{compare_1} show that the uncertainties of SOL2021-08-28T05 are too large to constrain the individual electron and flare parameters in this flare. The residuals and $\chi^{2}$ values of all simulation runs are consistent with the data (ranging from $\chi^{2}=0.5-1.7$) and hence, the electron directivity, high energy cutoff and local loop geometry cannot be constrained here. However, at this preliminary stage, the data is already consistent with the modeling outcomes; the expected 20-60 keV `bump' due to X-ray albedo is clearly visible. 

\section{Summary}\label{summ}
The measurement of HXR directivity from X-ray stereoscopy presents us with a prime opportunity for determining the unknown properties of solar flare acceleration, unavailable from X-ray spectra viewed from a single viewpoint, in particular:

\begin{enumerate}
    \item An electron angular (pitch-angle) distribution related to the acceleration process and coronal transport properties.
    \item The highest energy accelerated electrons related to the acceleration process and local plasma properties.
    \item The dominant magnetic field configuration via the property of `loop tilt'.
\end{enumerate}

The preliminary simulations outlined here can be compared directly with HXR spectroscopy results. In particular, the following observations will help to determine the electron directivity and other properties from HXR directivity:

\begin{enumerate}
\item If the HXR directivity at energies outside of $20-50$~keV (where the albedo component dominates), particularly at higher energies, is below 1, then the HXR directivity is suggestive of significant anisotropy (as shown in Figure \ref{fig3}).
\item How quickly HXR directivity falls to its lowest values at higher energies ($>70$~keV) helps to constrain the highest energy X-ray-producing electrons (as shown in Figure \ref{fig4}).
\item Large values of HXR directivity at its peak ($\approx 20-50$~keV) can be suggestive of high values of loop tilt, where the dominant loop direction is away from the local solar vertical direction (as shown in Figure \ref{fig5}).
\end{enumerate}

Our analysis shows the importance of accounting for the X-ray albedo component. For our preliminary stereoscopy study and comparison with simulation, we examined two flares: SOL2021-11-01T01 and SOL2021-08-28T05, observed by both SolO/STIX and Fermi/GBM. SOL2021-11-01T01 was used to check that a directivity of $\sim 1$ occurs when spacecraft view the flare at the same viewing angle, with the observation giving an average ratio of 1.05$\pm0.06$. SOL2021-08-28T05 did provide a directivity ratio greater than one, particularly over the dominant albedo energies of 20-50 keV, consistent with the albedo `bump' shown in the simulations. However, 
SOL2021-08-28T05 with its large uncertainties and small viewing angle difference of $\approx30^{\circ}$, is not ideal for directivity studies and electron and flare parameters could not be constrained in this observation.

As we move into the maximum of solar cycle 25, many more flares will be observed with multiple X-ray instruments and joint STIX-Fermi/GBM STIX-HXI, STIX-KONUS, and STIX-HEL1OS flare lists are being created. The different detector systems on board these missions have different strengths and difficulties. For example, Fermi/GBM has a relatively large effective area, but pileup issues are severe, particularly for large flares. For HXI, the $\sim$7 keV energy resolution is a limiting factor compared to the 1 keV value for STIX, but at the higher energies around $\sim$100 keV accurate measurements are potentially feasible. The PADRE CubeSat launched in late 2025 will make X-ray spectral measurements with the same detectors as STIX, likely providing the most accurate directivity measurements as systematic errors should be similar and therefore divide out when measuring the directivity. In any case, for each instrument the accuracy of the absolute flux calibration needs to be carefully established to be able to calculate meaningful error bars on the observed directivity. Before quantitative conclusions can be drawn, a study of the systematic uncertainties must be additionally performed by comparing spectra measured by the different detectors from similar viewing angles (i.e., within $\sim10$ degrees). As the orbit of SolO has a bias of spending relatively more time close to Earth or on the far side of the Sun, good observing times for directivity measurements with large angular separation are limited to two months around April and October, at least until after 2026 when SolO starts to leave the ecliptic. Nevertheless, we expect enough events to perform a statistical study. Finally, after preliminary comparisons, the next modelling steps will include the presence of multiple X-ray sources, the effects of more complex and realistic magnetic configurations and pitch-angle distributions derived from possible underlying acceleration mechanisms.

\section*{acknowledgments}
NLSJ gratefully acknowledges the current financial support from the Science and Technology Facilities Council (STFC) Grant ST/V000764/1. MS gratefully acknowledges financial support from a Northumbria University Research Development Fund (RDF) studentship.  The authors acknowledge IDL support provided by STFC. All authors are supported by an international team grant \href{https://teams.issibern.ch/solarflarexray/team/}{“Measuring Solar Flare HXR Directivity using Stereoscopic Observations with SolO/STIX and X-ray Instrumentation at Earth}” from the International Space Sciences Institute (ISSI) Bern, Switzerland. Solar Orbiter is a space mission of international collaboration between ESA and NASA, operated by ESA. The STIX instrument is an international collaboration between Switzerland, Poland, France, Czech Republic, Germany, Austria, Ireland, and Italy.
AFB, HC and SK are supported by the Swiss National Science Foundation Grant 200021L\_189180 for STIX.
AMV acknowledges the Austrian Science Fund (FWF): project no. I455-N. YS and FX acknowledge the National Natural Science Foundation of China (grant Nos. 11820101002, 12333010). The data that support the findings of this study are available from the corresponding author upon reasonable request.


\end{document}